\documentstyle[eqsecnum,aps,epsfig]{revtex}
\draft

\begin{document}
\newcommand{\bfk}{{\bf k}}
\newcommand{\bfn}{{\bf n}}
\newcommand{\bfp}{{\bf p}}
\newcommand{\bfq}{{\bf q}}
\newcommand{\bfr}{{\bf r}}
\newcommand{\bkappa}{\mbox{\boldmath $\kappa$}}
\newcommand{\bsigma}{\mbox{\boldmath $\sigma$}}
\newcommand{\btau}{\mbox{\boldmath $\tau$}}
\newcommand{\ii}{{\rm i}}
\newcommand{\bra}[1]{\langle #1 |}
\newcommand{\ket}[1]{| #1 \rangle}

\title{Distorted wave impulse approximation analysis for spin observables in
nucleon quasi-elastic scattering and enhancement of the spin-longitudinal 
response}
\author{Ken Kawahigashi\footnote{Electronic address: ken@info.kanagawa-u.ac.jp}}
\address{Department of Information Sciences, Kanagawa University, Hiratsuka 
259-1293, Japan}
\author{Kimiaki Nishida and Atsushi Itabashi}
\address{Institute of Physics, University of Tokyo, Komaba, Meguro-ku, Tokyo
153-8902, Japan }
\author{Munetake Ichimura\footnote{Electronic address: ichimura@k.hosei.ac.jp}}
\address{Faculty of Computer and Information Sciences, Hosei University, 
Koganei, Tokyo 184-8584, Japan}
\date{\today}

\maketitle

\begin{abstract}
\begin{quotation}
We present a formalism of distorted wave impulse approximation (DWIA) for 
analyzing spin observables in nucleon inelastic and charge exchange reactions
leading to the continuum. It utilizes response functions calculated by the 
continuum random phase approximation (RPA), which include the effective mass,
the spreading widths and the $\Delta$ degrees of freedom. The Fermi motion is
treated by the optimal factorization, and the non-locality of the 
nucleon-nucleon $t$-matrix by an averaged reaction plane approximation. By 
using the formalism we calculated the spin-longitudinal and the 
spin-transverse cross sections, $ID_q$ and $ID_p$, of $^{12}$C, $^{40}$Ca 
($\vec{p},\vec{n}$) at 494 and 346 MeV. The calculation reasonably reproduced
the observed $ID_q$, which is consistent with the predicted enhancement of the
spin-longitudinal response function $R_{\rm L}$. However, the observed $ID_p$
is much larger than the calculated one, which was consistent with neither
the predicted quenching nor the spin-transverse response function $R_{\rm T}
$ obtained by the ($e,e'$) scattering. The Landau-Migdal parameter $g'_ 
{N\Delta}$ for the $N\Delta$ transition interaction and the effective mass at
the nuclear center $m^*(r=0)$ are treated as adjustable parameters. The 
present analysis indicates that the smaller $g'_{N\Delta}(\approx 0.3)$ and 
$m^*(0) \approx 0.7 m$ are preferable. We also investigate the validity of the
plane wave impulse approximation (PWIA) with the effective nucleon number 
approximation for the absorption, by means of which $R_{\rm L}$ and $R_{\rm 
T}$ have conventionally been extracted.
\end{quotation}
\end{abstract}

\section{Introduction}

The study of nuclei by means of intermediate energy ($\sim$100 MeV--\ $\sim$1
GeV) proton beams has been very active since the end of the 1970's, and has 
led to various advances such as the discovery of the Gamov-Teller (GT) giant
resonance and finding the quenching of their strength. In the 1980's great 
progress was made in experimental facilities, which can afford to accelerate
polarized proton beams with intermediate energies and to measure the 
polarization of scattered protons and neutrons and so on. This made it 
possible to carry out complete measurements of ($\vec{p},\vec{p}'$) and ($\vec
{p},\vec{n}$) scattering, namely, measurement of the polarization $P$, the 
analyzing power $A_y$ and the polarization transfer coefficients $D_{ij}$. 
Theoretical investigations related to these experiments, such as search for 
the origin of the quenching of the GT strength and studies of the precursor 
phenomena of the pion condensation, have also been pursued vigorously \cite
{ref:OST92,ref:RS94}.

In the course of these activities, a very interesting prediction was presented
by Alberico {\it et al.}\cite{ref:AEM82} at the beginning of the 1980's. They
claimed that in the quasi-elastic region with fairly large momentum transfer
$q (\sim \mbox{1--3 fm}^{-1})$ the isovector spin-longitudinal response 
function $R_{\rm L}(q, \omega)$ is enhanced and softened while the isovector
spin-transverse response function $R_{\rm T}(q, \omega)$ is quenched and 
hardened, where $\omega$ is the transferred energy. The enhancement of $R_{\rm
L}$ is attributed to the collectivity induced by the one-pion exchange 
interaction and thereby understood as one of the precursor phenomena of pion
condensation. The quenching of $R_{\rm T}$, on the other hand, is due to the
combined effect of the repulsive short-range correlation and the one rho-meson
exchange interaction.

A great deal of experimental work has been done in order to explore this 
prediction. Measurements of the polarization transfer coefficients $D_{ij}$ 
were carried out at LAMPF for ($\vec{p},\vec{p}'$) at an incident energy of 
494 MeV, from which the ratios $R_{\rm L}/R_{\rm T}$ were extracted\cite
{ref:CA84,ref:RE86}. Surprisingly, the ratios were less than or equal to 
unity, which seriously contradicted the prediction  $R_{\rm L} \gg R_{\rm T}
$. However, the scattering ($\vec{p},\vec{p}'$) mixes the isoscalar and 
isovector contributions, and thus the estimation of the ratios was not 
conclusive.

Later complete measurements of the ($\vec{p},\vec{n}$) reaction, which focused
exclusively on the isovector contribution, were carried out at LAMPF\cite 
{ref:MC92,ref:CHEN93,ref:TA94} and at RCNP\cite{ref:WA99}. The ratios obtained
were still less than unity. From these results, it was concluded\cite 
{ref:BFS93,ref:BBLW95} that there is no enhancement of $R_{\rm L}$, namely no
collective enhancement of the pionic modes, which was interpreted as evidence
against the pion excess in the nucleus. With help of sum rules, Koltun\cite 
{ref:KO98} analyzed the data by means of a correlated nuclear theory, in which
the two-nucleon correlation is dominating, and concluded that there is no 
disagreement between the data and the model for $R_ {\rm L}$, but $R_{\rm T}
$ is not explained. He claimed that there is no collective enhancement of the
pionic modes. However, we believe that there are many questions to be solved
before such conclusions are reached.

The experimental evaluation of the ratio was based on the plane wave impulse
approximation (PWIA) with effective nucleon number approximation for 
absorption where the number of participant nucleons is estimated by a simple
eikonal approximation. Evidently we should take into account the effects of 
distortion more accurately. Since this causes the nuclear responses to be of
surface nature, the semi-infinite slab model\cite{ref:BS82,ref:ETB85}, the 
surface response model\cite {ref:Al86,ref:Al88} and others, were introduced.
The distorted wave Born approximation (DWBA) analysis of the continuum spectra
of the intermediate energy ($p, n$) reaction was carried out by Izumoto {\it
et al.}\cite {ref:IIKS82} where the response functions were calculated by the
local density approximation. The calculation in the distorted wave impulse 
approximation (DWIA) for the spin-longitudinal and -transverse cross sections,
$ID_q$ and $ID_p$, in the continuum was first performed by Ichimura {\it et 
al.}\cite {ref:IKJG89} for ($\vec{p}, \vec{p}'$). We utilized the response 
functions calculated by the continuum random phase approximation (continuum 
RPA) with orthogonality condition\cite{ref:Izu83} including the $\Delta$ 
degrees of freedom. We there employed the optimal factorization 
prescription\cite {ref:PTTW84,ref:Gu86,ref:ZMW87} to deal with the Fermi 
motion of the struck nucleon, restricted the spin dependence of the 
nucleon-nucleon ($NN$) $t$-matrix to the terms with $(\bsigma_0\cdot\hat{\bf
q})(\bsigma_1\cdot\hat {\bf q})$, and $(\bsigma_0\times\hat{\bf q})\cdot
(\bsigma_1\times\hat{\bf q}) $ and neglected the interference between the 
different spin-dependent terms of the $NN$ $t$-matrix. De Pace\cite{ref:DP95}
treated the absorption by the Glauber approximation and calculated the 
spin-longitudinal and -transverse cross section up to the two-step processes.
He also used the same assumption for the $NN$ $t$-matrix. Recently, Kim {\it
et al.}\cite{ref:KKSU00} also developed a DWIA formalism for calculating the
spin observables in a form of inhomogeneous coupled-channel integral equations
in the Tamm-Dancoff approximation. They also used the same assumption for the
$NN$ interaction and further neglected the spin orbit force in the optical 
potential. Noting this situation, we definitely need a more reliable method 
for analyzing the spin observables as well as the inclusive cross sections in
the continuum.

In this paper, we develop a DWIA formalism for nucleon-nucleus ($NA$) 
scattering at the intermediate energy leading to the continuum, with the 
response functions non-diagonal with respect to the momentum transfer as well
as the spin directions. This method does not require the above restriction on
the $NN$ $t$-matrix and can handle the interference between the different 
spin-dependent terms. Thus it is much more reliable for calculating the spin
observables such as $P, A_y$ and $D_{ij}$.

The original prediction of Alberico {\it et al.}\cite{ref:AEM82} was based on
the theoretical framework of (1) the Fermi gas model, (2) RPA, or more 
precisely the ring approximation, with the one-pion + one-rho-meson exchange
interactions + the contact interaction specified by the Landau-Migdal 
parameters, $g'$'s ($\pi+\rho+g'$ model), and (3) the universality ansatz, 
$g'_{NN}=g'_{N\Delta}=g'_{\Delta\Delta}$. A number of improvements have been
made to various aspects of the method, such as use of the continuum RPA to 
treat the nuclear size effects\cite{ref:SSA88,ref:IKJG89}, and removal of the
universality ansatz\cite{ref:SSIT86,ref:NI95}. In the previous continuum RPA
calculations, the mean field was assumed to be local and the spreading width
of the particle was treated by the complex potential, but that of the hole was
neglected. In this paper we take account of the non-locality of the mean field
by the radial-dependent effective mass, $m^*(r)$, and the spreading width of
the hole by the complex binding energy.

Special importance has been put on the choice of the Landau-Migdal parameters. 
The position of the GT resonance gives severe restriction on the value of $g'_
{NN}$ ($\approx 0.6-0.7$). Effects of $\Delta$-isobar are very sensitive to 
the value of $g'_{N\Delta}$\cite{ref:SSIT86,ref:NI95}, which is also crucial
for the cause of the quenching of the GT strength\cite{ref:SS99}. In the 
present analysis of the quasi-elastic ($\vec{p}, \vec{n}$) reaction, we treat
$g'_{N\Delta}$ and the effective mass at the center of the nucleus $m^*(r=0)
$ as the adjustable parameters, and thus try to evaluate the value of $g'_
{N\Delta}$. This is one of the main aims of the present paper.

This paper is organized as follows. In Sec.~\ref{sec:form} we summarize the 
general formulas for the cross sections and the spin observables of the ($\vec
{N},\vec{N}'$) reactions, with special attention to the polarized cross 
sections $ID_i$. In Sec.~\ref{sec:PWIA}, the PWIA formalism with the optimal
factorization and the effective nucleon number approximation is briefly 
reviewed because it has been used to extract $R_{\rm L}$ and $R_{\rm T}$ up 
to now.

In Sec.~\ref{sec:DWIA} we present the DWIA formalism for the intermediate 
energy ($\vec{N}, \vec{N}'$) reactions to the continuum in cooperation with 
the response functions, again utilizing the optimal factorization. Since we 
treat the finite nucleus, calculations are carried out in the angular momentum
representation, details of which are presented in Sec.~\ref{sec:ang}. We 
describe a way of calculating the response functions non-diagonal in the 
coordinate space as well as the spin space, which involves the 
radial-dependent effective mass and the spreading widths of both the particle
and the hole in Sec.~\ref{sec:crpa}.

In Sec.~\ref{sec:num}, we perform numerical analysis for the reactions $^{12}
$C, $^{40}$Ca($\vec{p},\vec{n}$) at 494 MeV and at 346 MeV. We see that the 
spin-longitudinal cross sections are reasonably well reproduced. This is 
consistent with the predicted enhancement of $R_{\rm L}$ though the ratio $R_
{\rm L}/R_{\rm T}$ is less than unity. However, the spin-transverse cross 
sections are very much underestimated. The contradiction regarding the ratio
seems to come from the large difference between the experiments and the 
theories in $ID_p$. This confirms the reported conclusion in the experimental
papers\cite{ref:TA94,ref:WA99}.

In Sec.~\ref{sec:eik} we test the reliability of the conventional method for
extracting the response functions by comparing the results of DWIA and PWIA 
with the effective nucleon number. Sec.~\ref{sec:dis} is devoted to some other
questions such as the effects of the spin-orbit force and the ambiguity of the
optical potential. Sec.~\ref {sec:sum} consists of a summary and conclusion.

\section{General formulas for cross section and spin observables}
\label{sec:form}

We consider the nucleon-nucleus $(NA)$ inelastic and charge-exchange reactions
to the continuum. First we summarize the general formulas for the double 
differential cross section and the spin observables of a polarized nucleon 
scattering off a nucleus,
\begin{equation}
\vec{N} + A  \rightarrow  \vec{N}' + B. 
\label{reaction}
\end{equation}

In the $NA$ center of mass (c.m.) system, we denote the scattering angle by 
$\theta_{\rm cm}$, and the momenta of the incident and outgoing nucleons by 
$\bfk_i$ and $\bfk_f$, respectively. The momentum transfer to the scattering
nucleon is given by \begin{equation} \bfq = \bfk_{f} - \bfk_{i}, \end
{equation} and the energy transfer to the target is written as \begin
{equation} \omega_{\rm cm} = E_{N}(k_{i}) - E_{N'}(k_{f}), \end{equation} 
where $E_{\alpha}(k) = \sqrt{k^{2} + m_{\alpha}^{2}}$\ \ is the energy of a 
particle $\alpha$ with mass $m_{\alpha}$. We use the unit system $\hbar=1$ and
$c=1$ throughout this paper.

As the coordinate system, we use either
\begin{equation}
\hat{\bf z} = \frac{\bfk_{i}}{|\bfk_{i}|},\ \ \ \ 
\hat{\bf y} = \frac{\bfk_{i} \times \bfk_{f}}{|\bfk_{i} \times \bfk_{f}|},
\ \ \ \ \hat{\bf x} = {\hat{\bf y} \times \hat{\bf z}}
\end{equation}
or
\begin{equation}
\hat{\bfq} = \frac{\bfq}{\mid \bfq \mid}, \ \ \ \ \hat{\bfn} = \frac{\bfk_{i}
\times \bfk_{f}} {\mid \bfk_{i} \times \bfk_{f} \mid}, \ \ \ \ 
\hat{\bfp} = \hat{\bfq} \times \hat{\bfn}.   
\label{eq:direc-cm}
\end{equation}
These are called the $x, y, z$ and $q, n, p$ directions, respectively.

The unpolarized double differential cross section is expressed by the $NA$ 
$t$-matrix $T$ as
\begin{eqnarray}
I(\theta_{\rm cm},\omega_{\rm cm}) &\equiv&\frac{d^2 \sigma}{d\Omega_{\rm cm}
d\omega_{\rm cm}} \nonumber \\
&=& \frac{1}{2(2J_A+1)}\frac{\mu_{i}\mu_{f}}{(2\pi)^{2}}\frac{k_{f}}{k_{i}}
\sum_{m_{s_f} m_{s_i}}\sum_n {\sum_{n_0}}' |\bra{\bfk_f m_{s_f} \Psi_n} T
\ket{\bfk_i m_{s_i} \Psi_{n_0}}|^2 \nonumber \\
& & \times\delta\left(\omega_{\rm cm} - (E_{B}(k_{f}) - E_{A}(k_{i}))\right),
\label{eq:Icm}
\end{eqnarray}
where $m_{s_i}$($m_{s_f}$) is the spin projection of the incident (outgoing)
nucleon and $J_A$ is the target spin, and $\mu_{i}$ and $\mu_{f}$ are the
relativistic reduced energies
\begin{equation}
\mu_{i} = \frac{E_{N}(k_{i})E_{A}(k_{i})}{\sqrt{s_{NA}}},\ \ \ \ 
\mu_{f} = \frac{E_{N'}(k_{f})E_{B}(k_{f})}{\sqrt{s_{NA}}}
\end{equation}
with $s_{NA}=\left(E_{N}(k_i) + E_{A}(k_i)\right)^2$. The wave functions
$\Psi_{n_0}$ and $\Psi_{n}$ are intrinsic states of the target $A$ and of
the residual nucleus $B$, respectively. They are governed by the intrinsic
Hamiltonian $H_{A}$ of the $A$-body system as
\begin{equation}
H_{A}\Psi_{n}={\cal E}_{n}\Psi_{n}
\end{equation}
where $H_{A}$ includes the mass terms and thus ${\cal E}_{n}$ means the 
invariant mass of the $A$-body systems. Note that the final states $\Psi_n$ 
are mostly unbound. In the summation $\Sigma'_{n_0}$, $n_0$ runs only over the
degenerate ground states of the target $A$.

To extract information of intrinsic states separately, we introduce the 
intrinsic energy transfer $\omega$ and rewrite the cross section of Eq.~(\ref
{eq:Icm}) as
\begin{equation}
I(\theta_{\rm cm},\omega_{\rm cm}) = \frac{K}{2(2J_A+1)}{\rm Tr}\left[{\rm 
Tr'}\left[TT^{\dagger}\right]\right], \label{eq:Icm2}
\end{equation}
with
\begin{equation}
{\rm Tr'}\left[TT^{\dagger}\right]=\sum_n{\sum_{n_0}}' \bra {\Psi_n} T \ket
{\Psi_{n_0}} \bra{\Psi_{n_0}} T^{\dagger} \ket{\Psi_n} \delta\left(\omega - 
({\cal E}_{n} - {\cal E}_{n_0})\right),
\end{equation}
where Tr means the trace of the spin states of the incident and exit nucleons.
The kinematical factor $K$ is given by
\begin{equation}
K=\frac{\mu_{i}\mu_{f}}{(2\pi)^{2}}\frac{k_{f}}{k_{i}}
\frac{d\omega}{d\omega_{\rm cm}}=\frac{\mu_{i}\mu_{f}}
{(2\pi)^{2}}\frac{k_{f}}{k_{i}}\frac{\sqrt{s_{NA}}}{{\cal E}_{n}}
\label{eq:kfacp}
\end{equation}
using the relation $\frac{d\omega}{d\omega_{\rm cm}}=\frac{\sqrt{s_{NA}}}
{{\cal E}_{n}}$ \cite{ref:NAI99}. We note that the present $K$ is equal to $2
(2J_A+1)CK$ of Ref.~\cite{ref:WA99} , $2(2J_A+1)K\frac{d\omega}{d\omega_{\rm
cm}}$ of Ref.~\cite{ref:IK92} and $K\frac{d\omega}{d\omega_{\rm cm}}$ of 
Ref.~\cite{ref:NAI99}.

We represent the Pauli spin operators in the $i$-direction of the $k$-th 
nucleon by $\sigma_{ki}$. To unify the notation we also introduce $\sigma_{k0}
(={\bf 1}_k)$ for the unit spin matrix of the $k$-th nucleon. The nucleon 
number $k=0$ denotes the incident or exit nucleon. Then the polarization, the
analyzing power and the polarization transfer coefficients are given 
respectively by
\[
P_y = \frac{{\rm TrTr'}[TT^{\dag}\sigma_{0y}]} {{\rm TrTr'}[TT^{\dag}]},
\ \ \ \ A_y = \frac{{\rm TrTr'}[T\sigma_{0y} T^{\dag}]} {{\rm TrTr'}
[TT^{\dag}]},
\]
\begin{equation}
D_{ij} = \frac{{\rm TrTr'} [T\sigma_{0i} T^{\dag} \sigma_{0j} ]} {{\rm TrTr'}
[TT^{\dag}]}, \label{eq:SpOb}
\end{equation}
where $i,j=x,y,z$ or $q,n,p$.

The DWIA calculation is usually carried out in the $[\hat{\bf x},\hat{\bf y}
,\hat{\bf z}]$ frame, while the $[\hat{\bfq},\hat{\bfn},\hat{\bfp}]$ frame is
sometimes more convenient for the theoretical analysis. The relation between
$D_{ij}\ (i,j=q,n,p)$ and $D_{ij}\ (i,j=x,y,z)$ is given by
\[
D_{nn}=D_{yy},
\]
\begin{equation}
\left( \begin{array}{cc} D_{pp} & D_{pq} \\ D_{qp} & D_{qq} \end{array}
\right)=\left( \begin{array}{cc} \cos\theta_{\rm p} & \sin\theta_{\rm p} \\
-\sin\theta_{\rm p} & \cos\theta_{\rm p} \end{array} \right)\left( 
\begin{array}{cc} D_{zz} & D_{zx} \\ D_{xz} & D_{xx} \end{array} \right)
\left( \begin{array}{cc} \cos\theta_{\rm p} & -\sin\theta_{\rm p} \\ 
\sin\theta_{\rm p} & \cos\theta_{\rm p} \end{array} \right),
\label{eq:x2q} 
\end{equation}
where $\theta_{\rm p}$ is the angle between $\hat{\bf p}$ and $\hat{\bf z}$.

To extract nuclear responses, Bleszynski {\it et al.}\cite{ref:BL82} 
decomposed the $NA$ $t$-matrix in the $[\hat{\bfq},\hat{\bfn},\hat{\bfp}]$ 
frame as
\begin{equation}
T = T_{0}\sigma_{00} + T_{n}\sigma_{0n} + T_{q} \sigma_{0q}
+ T_{p}\sigma_{0p}
\label{eq:Ti}
\end{equation}
and introduced the polarized cross sections $ID_i$, which extract $T_{i}$
exclusively as
\begin{eqnarray}
ID_{0} &=& \frac{I}{4}[1 + D_{nn} + D_{qq} + D_{pp}] =  \frac{K}{2J_A+1}
{\rm Tr'}[T_{0}T_{0}^{\dag}], \nonumber \\
ID_{n} &=& \frac{I}{4}[1+ D_{nn}- D_{qq} - D_{pp}] =  \frac{K}{2J_A+1}
{\rm Tr'}[T_{n}T_{n}^{\dag}],
\nonumber \\
ID_{q} &=& \frac{I}{4}[1 - D_{nn} + D_{qq} - D_{pp}] =  \frac{K}{2J_A+1}
{\rm Tr'}[T_{q}T_{q}^{\dag}], \nonumber \\
ID_{p} &=& \frac{I}{4}[1 - D_{nn} - D_{qq} + D_{pp}] =  \frac{K}{2J_A+1}
{\rm Tr'}[T_{p}T_{p}^{\dag}].
\label{eq:IDi}
\end{eqnarray}
The unpolarized cross section $I$ is expressed as
\begin{equation}
I=ID_0+ID_n+ID_q+ID_p.
\end{equation}

In the $NA$ laboratory system we denote the angle, the momenta and the energy
transfer corresponding to $\theta_{\rm cm},\bfk_i,\bfk_f$ and $\omega_{\rm cm}
$ by $\theta_{\rm lab}$, ${\bf K}_i, {\bf K}_f$ and $\omega_{\rm lab}$, 
respectively. The unit vectors
\begin{eqnarray}
\hat{\bf N} = \hat{\bf N}' =
\frac{{\bf K}_{i} \times {\bf K}_{f}}{|{\bf K}_{i} \times {\bf K}_{f}|},
 \ \ \ \hat{\bf L} &=& \hat{\bf K}_{i},
 \ \ \ \hat{\bf S} = \hat{\bf N} \times \hat{\bf L},  \nonumber \\
\hat{\bf L}' &=& \hat{\bf K}_{f},
\ \ \ \hat{\bf S}' = \hat{\bf N}' \times \hat{\bf L}'
\label{direc-lab}
\end{eqnarray}
are usually used to specify the directions, and are denoted by $N, L, S, N',
 L'$ and $S'$, respectively.

In this system, the unpolarized cross section is given by
\begin{equation}
I_{\rm lab}(\theta_{\rm lab},\omega_{\rm lab}) = \frac{d^2 \sigma}{d\Omega_
{\rm lab} d\omega_{\rm lab}} = \frac{K_f}{k_f}I(\theta_{\rm cm},\omega_{\rm 
cm}),
\end{equation}
due to the relation\cite{ref:HA63}
\begin{equation}
\frac{d\Omega_{\rm cm} d\omega_{\rm cm}}{d\Omega_{\rm lab} d\omega_{\rm lab}}
=\frac{\sin\theta_{\rm cm}}{\sin\theta_{\rm lab}}=\frac{K_f}{k_f},
\label{eq:cm2lab}
\end{equation} 
and the observed polarization transfer coefficients $D_{ij} (i=S,N,L; 
j=S',N',L')$ are obtained from the calculated $D_{ij} (i,j=x,y,z)$ as
\begin{eqnarray}
D_{NN'} &=& D_{yy} \nonumber \\
\left(\begin{array}{cc} D_{LL'} & 
D_{LS'} \\ D_{SL'} & D_{SS'} \\ \end{array}\right) &=& \left(\begin{array}{cc}
D_{zz} & D_{zx} \\ D_{xz} & D_{xx} \\ \end{array}\right) \left(\begin{array}
{cc} \cos\alpha_1  & -\sin\alpha_1 \\ \sin\alpha_1  & \cos\alpha_1 \\ \end
{array}\right) \label{eq:x2s}
\end{eqnarray}
with $\alpha_1=\theta_{\rm lab}
+\Omega$, where $\Omega$ is the relativistic spin rotation angle\cite
{ref:IK92}.

By use of Eqs.~(\ref{eq:x2q}), (\ref{eq:IDi}) and (\ref{eq:x2s}), $D_i$'s are
obtained from $D_{ij}$'s in the $NA$ laboratory frame as\cite{ref:IK92}
\begin{eqnarray}
D_{0} &=& \frac{1}{4}[1 + D_{NN} + (D_{SS'}+D_{LL'})\cos\alpha_1 \nonumber \\
& & \hspace*{2.1cm} \mbox{}+ (D_{SL'}-D_{LS'})\sin\alpha_1], \nonumber \\
D_{n} &=& \frac{1}{4}[1 + D_{NN} - (D_{SS'}+D_{LL'}) \cos\alpha_1 \nonumber \\
& & \hspace*{2.1cm} \mbox{}- (D_{SL'}-D_{LS'}) \sin\alpha_1], \nonumber \\
D_{q} &=& \frac{1}{4}[1 - D_{NN} + (D_{SS'}-D_{LL'})\cos\alpha_2 \nonumber \\
& & \hspace*{2.1cm} \mbox{}- (D_{LS'}+D_{SL'})\sin\alpha_2], \nonumber \\
D_{p} &=& \frac{1}{4}[1 - D_{NN} - (D_{SS'}-D_{LL'})\cos\alpha_2 \nonumber \\
& & \hspace*{2.1cm} \mbox{}+ (D_{LS'}+D_{SL'})\sin\alpha_2]
\end{eqnarray}
with $\alpha_2=2\theta_{p}-\alpha_1$.

\section{PWIA formalism}
\label{sec:PWIA}

Before presenting the DWIA formalism, we briefly review the PWIA 
formalism\cite {ref:IK92}, which has conventionally been used in the analysis
of quasi-elastic scattering to extract the spin response functions\cite 
{ref:CA84,ref:RE86,ref:MC92,ref:CHEN93,ref:TA94,ref:WA99}. To avoid confusion,
we suppress the isospin and $\Delta$ degrees of freedom until Sec.~\ref
{sec:isodel}, where we discuss these degrees of freedom in detail.

\subsection{PWIA $t$-matrix} 

The PWIA $t$-matrix in the $NA$ c.m.\ system is written as
\begin{equation}
T_{nn_0}^{\rm PW}(\bfk_f, \bfk_i) = \bra{\Psi_n} \sum_{k=1}^A t_k(\bfk_f, 
\bfk_i) \ket{\Psi_{n_0}} \label{eq:tpw},
\end{equation}
where $t_k(\bfk_f, \bfk_i)$ is the $NN$ scattering $t$-matrix between the 
incident nucleon and the $k$-th nucleon in the nuclei.

To avoid the difficulty of Fermi momentum integration, the optimal 
factorization approximation\cite{ref:PTTW84,ref:Gu86,ref:ZMW87,ref:IK92} is 
often used, where $t_k(\bfk_f, \bfk_i)$ is replaced by the $NN$ $t$-matrix in
the optimal frame, $t^{\eta}_k(\bfk_f,\bfk_i)$, which is written as
\begin{eqnarray}
t^{\eta}_k(\bfk_f,\bfk_i)&=&\bra{\bfk_f, \tilde{\bfp}'} t \ket{\bfk_i, \tilde
{\bfp}} = \sum_{ab}t^{\eta}_{ab}(\bfk_f,\bfk_i)\sigma_{0a}\sigma_{kb} 
\nonumber \\
   &=& A^{\eta}    \sigma_{00}\sigma_{k0} + B^{\eta}    \sigma_{0n}\sigma_{kn}
   + C^{\eta}_{1}\sigma_{0n}\sigma_{k0} + C^{\eta}_{2}\sigma_{00}\sigma_{kn}
   \nonumber \\
   &+& D^{\eta}_{1}\sigma_{0p}\sigma_{kq} + D^{\eta}_{2}\sigma_{0q}\sigma_{kp}
   + E^{\eta}    \sigma_{0q}\sigma_{kq} + F^{\eta}    \sigma_{0p}\sigma_{kp}
   .
   \label{eq:topt}
\end{eqnarray}
The optimal momenta $\tilde{\bfp}$ and $\tilde{\bfp}'$ of the struck nucleon
in the nucleus are given by
\begin{equation}
\tilde{\bfp} = (\frac{1}{2} - \eta)\bfq-\frac{\bfk_i+\bfk_f}{2A},\ \ \tilde
{\bfp}'=\tilde{\bfp}-\bfq, \label{eq:tildep}
\end{equation}
respectively, and the parameter $\eta$ is determined by the on-shell condition
$E_N(\bfk_i)+E_N(\tilde{\bfp})=E_N(\bfk_f)+E_N(\tilde{\bfp}')$.

This $NN$ $t$-matrix in the optimal frame is obtained from the observed $NN$
$t$-matrix in the c.m.\ frame,
\begin{eqnarray}
t^{\rm cm}(\bkappa',\bkappa) &=& \bra{\bkappa', -\bkappa'} t \ket{\bkappa, 
-\bkappa} =\sum_{a',b'}t^{\rm cm}_{a'b'}(\bkappa',\bkappa)\sigma_{0a'}\sigma_
{1b'} \nonumber \\
&=& A'\sigma_{00}\sigma_{10} + B'\sigma_{0n_{\rm c}}\sigma_{1n_{\rm c}} + C'
(\sigma_{0n_{\rm c}}\sigma_{10} + \sigma_{00}\sigma_{1n_{\rm c}}) + E'\sigma_
{0q_{\rm c}}\sigma_{1q_{\rm c}} + F'\sigma_{0p_{\rm c}}\sigma_{1p_{\rm c}},
\end{eqnarray}
where $\bkappa$ ($\bkappa'$) is the initial (final) relative momentum in the
$NN$ c.m.\  scattering, which is determined from $\bfk_i$ ($\bfk_f$) and 
$\tilde{\bfp}$ ($\tilde{\bfp}'$). Here the coordinate system is determined by
\begin{equation}
\hat{\bfq}_{\rm c} = \frac{\bfq_{\rm c}} {|\bfq_{\rm c}|}, \ \ \
\hat{\bfn}_{\rm c} = \frac{\bkappa \times \bkappa'} {\mid \bkappa \times 
\bkappa' \mid} = \hat{\bfn}, \ \ \
\hat{\bfp}_{\rm c} = \hat{\bfq}_{\rm c} \times \hat{\bfn}_{\rm c}
\end{equation}
with $\bfq_{\rm c} = \bkappa'-\bkappa$. The relation between $t^{\eta}
(\bfk_f,\bfk_i)$ and $t^{\rm cm}(\bkappa',\bkappa)$ is given by
\begin{equation}
t^{\eta}(\bfk_f,\bfk_i) = J_{\eta}(\bfk_f,\bfk_i)R^{\rm \ell}_{\rm spin}t^{\rm
cm}(\bkappa',\bkappa)R^{\rm r}_{\rm spin}.
\end{equation}
The M\"{o}ller factor $J_{\eta}$ is given by
\begin{equation}
J_{\eta}(\bfk_f,\bfk_i)=\frac{{E_N(\bkappa)}^2} {\sqrt{E_N(\bfk_i)E_N(\bfk_f)
E_N(\tilde{\bfp})E_N(\tilde{\bfp}')}},
\end{equation}
where we neglect the mass difference between the proton and the neutron. The
relativistic spin rotations $R^{\rm \ell}_{\rm spin}$ and $R^{\rm r}_{\rm 
spin}$ are given in Eq.~(3.29) of Ref.~\cite{ref:IK92}, and the relation 
between $t^{\eta}_{ab}$ and $t^{\rm cm}_{a'b'}$ is explicitly given in
Eq.~(3.34) of Ref.~\cite{ref:IK92}.

The approximation $\bfq \approx \bfq_{\rm c}$ holds\cite{ref:IK92} for the 
reactions we are interested in. Noting that $\hat{\bfn}_{\rm c} = \hat{\bfn}
$, we can identify the coordinate system $[\hat{\bfq}, \hat{\bfn}, \hat{\bfp}]
$ with $[\hat{\bfq}_{\rm c}, \hat{\bfn}_{\rm c}, \hat{\bfp}_{\rm c}]$. We also
showed\ \cite{ref:IK92} that the approximation $R^{\rm \ell(r)}_{\rm spin} 
\approx 1$ holds. Therefore we can safely write
\begin{equation}
t^{\eta}_{ab}(\bfk_f,\bfk_i) = J_{\eta}(\bfk_f,\bfk_i)t^{\rm cm}_{ab}
(\bkappa',\bkappa),
\end{equation}
namely,
\begin{eqnarray}
& & A^{\eta}=J_{\eta}A',\ B^{\eta}=J_{\eta}B',\ C^{\eta}_1=C^{\eta}_2=J_{\eta}
C',\nonumber \\
& & D^{\eta}_1=D^{\eta}_2=0,\ E^{\eta}=J_{\eta}E',\ F^{\eta}=J_{\eta}F'.
\label{eq:teta}
\end{eqnarray}

From Eqs.~(\ref{eq:tpw}) and (\ref{eq:topt}), the PWIA $t$-matrix with the 
optimal factorization is given by
\begin{equation}
T_{nn_0}^{\rm PW}(\bfk_f, \bfk_i)=\sum_{a}\sum_{b}t_{ab}^{\eta}(\bfk_f, 
\bfk_i)F_{b}^{nn_0}(\bfq) \sigma_{0a}
\label{eq:tpwopt}
\end{equation}
with the transition form factor
\begin{equation}
F_{b}^{nn_0}(\bfq)=\bra{\Psi_n}\sum_{k=1}^A \sigma_{kb}{\rm e}^{-i\bfq\cdot 
\bfr_{k}} \delta\left(\sum_{k'=1}^A\bfr_{k'}\right)\ket{\Psi_{n_0}}.
\label{eq:trff}
\end{equation}

\subsection{Response Functions}
\label{subsec:resp}

From Eq.~(\ref{eq:IDi}), the polarized cross sections $I_{\rm lab}D_i$ in PWIA
are expressed as
\begin{equation}
I_{\rm lab}^{\rm PW}D_i^{\rm PW}=K_{\rm lab} \sum_{ab}t_{ia}^{\eta}(\bfk_f, 
\bfk_i) R_{ba}(q, \omega)t_{ib}^{\eta *}(\bfk_f, \bfk_i),
\label{eq:IDiPW}
\end{equation}
where $K_{\rm lab}=(K_f/k_f) K$ (see Eq.~(\ref{eq:cm2lab})) and $R_{ba}(q, 
\omega)$ are the response functions
\begin{equation}
R_{ba}(q,\omega) = \frac{1}{2J_A+1}\sum_n {\sum_{n_0}}' F_{a}^{nn_0}(\bfq)
F_{b}^{nn_0*}(\bfq)\delta(\omega - ({\cal E}_n - {\cal E}_{n_0})),
\label{eq:Rint}
\end{equation}
which are determined solely by the nuclear intrinsic states. Note that $R_{ba}
$ depends on only the magnitude $q=|\bfq|$, because of the unpolarized target,
i.e.~the presence of the summation $\sum'_{n_0}$.

For comparison with the results of different nuclei, the {\it normalized} 
response functions
\begin{equation}
\hat{R}_{ba}(q, \omega)=\frac{1}{A}R_{ba}(q, \omega),
\label{eq:Rba}
\end{equation}
whose spin diagonal parts satisfy the relation
\begin{equation}
\int \hat{R}_{aa}(q, \omega) d\omega \rightarrow 1 \ \ \mbox{for}\ \ q 
\rightarrow \infty,
\label{eq:Rlim}
\end{equation}
are more convenient.

Noting Eq.~(\ref{eq:teta}), $I^{\rm PW}_{\rm lab}D^{\rm PW}_q$ and $I^{\rm PW}
_{\rm lab}D^{\rm PW}_p$ are simply written, using the normalized 
spin-longitudinal and spin-transverse response functions $\hat{R}_{\rm L}$ and
$\hat{R}_{\rm T}$ as
\begin{eqnarray}
I_{\rm lab}^{\rm PW}D_q^{\rm PW}&=&K_{\rm lab} A |E^{\eta}(\bfk_f, \bfk_i)|^2
\hat{R}_{\rm L}(q,\omega), \nonumber \\
I_{\rm lab}^{\rm PW}D_p^{\rm PW}&=&K_{\rm lab} A |F^{\eta}(\bfk_f, \bfk_i)|^2
\hat{R}_{\rm T}(q,\omega)
\label{eq:IDQPPW}
\end{eqnarray}
where
\begin{eqnarray}
\hat{R}_{\rm L}(q,\omega)&=&\hat{R}_{qq}(q,\omega), \nonumber \\
\hat{R}_{\rm T}(q,\omega)&=&\frac{1}{2} \left(\hat{R}_{pp}(q,\omega)
+\hat{R}_{nn}(q,\omega)\right) = \hat{R}_{pp}(q,\omega).
\label{eq:Rhat}
\end{eqnarray}

\subsection{Effective nucleon number approximation for absorption}

In practical analysis we must include the effects of the absorption, which are
usually treated by the effective nucleon number $A_{\rm eff}$ estimated by the
simple eikonal approximation as
\begin{equation}
A_{\rm eff}=\int_0^{\infty} 2\pi b n(b) \exp\left[-n(b)\tilde{\sigma}_{NN}
\right] db
\label{eq:Aeff}
\end{equation}
with $n(b)=\int_{-\infty}^{\infty} dz \rho_A(\sqrt{z^2+b^2})$ where $b$ is the
impact parameter, $\tilde{\sigma}_{NN}$ is the total $NN$ cross section in the
nuclear medium, and $\rho_A$ is the nuclear density.

Then $I_{\rm lab}D_i$'s in PWIA with absorption, $\tilde{I}_{\rm lab}\tilde{D}
_i$, are given with $A_{\rm eff}$ by
\begin{equation}
\tilde{I}_{\rm lab}\tilde{D}_i=K_{\rm lab} A_{\rm eff} \sum_{ab}t_{ia}^{\eta}
(\bfk_f, \bfk_i) \hat{R}_{ba}(q,\omega)t_{ib}^{\eta *}(\bfk_f, \bfk_i),
\end{equation}
and especially
\begin{eqnarray}
\tilde{I}_{\rm lab}\tilde{D}_q&=&K_{\rm lab} A_{\rm eff} |E^{\eta}(\bfk_f, 
\bfk_i)|^2 \hat{R}_{\rm L}(q,\omega), \nonumber \\
\tilde{I}_{\rm lab}\tilde{D}_p&=&K_{\rm lab} A_{\rm eff} |F^{\eta}(\bfk_f, 
\bfk_i)|^2 \hat{R}_{\rm T}(q,\omega).
\label{eq:IDIwA}
\end{eqnarray}
This prescription is called the effective nucleon number approximation.

\subsection{Isospin and $\Delta$ degrees of freedom}
\label{sec:isodel}

Now let us take account of the isospin and the $\Delta$ isobar degrees of 
freedom. Following Ref.\cite{ref:NI95} with some modifications, we introduce
the unified notation of spin and isospin operators, $\sigma^{\alpha}_{ka}$ and
$\tau^{\alpha}_{k\kappa}$, of the $k$-th particle ($N$ or $\Delta$) as
\begin{equation}
\sigma^{\alpha}_{ka}=\left\{
\begin{array}{ll}
\sigma_{ka} & \mbox{for $\alpha=N$} \\
S_{ka}^{\dag} & \mbox{for $\alpha=\Delta$} \\
S_{ka} & \mbox{for $\alpha=\Delta^*$}
\end{array} \right.
\end{equation}
and
\begin{equation}
\tau^{\alpha}_{k\kappa}=\left\{
\begin{array}{ll}
\tau_{k\kappa} & \mbox{for $\alpha=N$} \\
(T_{k}^{\dag})_{\kappa} & \mbox{for $\alpha=\Delta$} \\
T_{k\kappa} & \mbox{for $\alpha=\Delta^*$}
\end{array} \right.
\label{eq:tau}
\end{equation}
with $\kappa={\rm s},0,+,-$ where
\begin{equation}
\tau_{k \rm s}={\bf 1}_{k},\ \ \tau_{k \pm}=\frac{\tau_{kx}\pm \ii\tau_{ky}} 
{\sqrt{2}},\ \ \tau_{k0}=\tau_{kz}
\end{equation}
and ${\bf S}_k$ and ${\bf T}_k$ are the standard spin and isospin transition
operators from $N$ to $\Delta$.

We extend the $NN$ $t$-matrix of Eq.~(\ref{eq:topt}) to that which includes 
the isospin and the $N\Delta$ channel in a restricted form as
\begin{eqnarray}
t^{\eta}_k&=&t^{\eta (0)}_k+t^{\eta (1)}_k\btau_{0}\cdot\btau_{k}+t^{\eta 
(\Delta)}_k \btau_{0}\cdot {\bf T}_{k}+t^{\eta (\Delta^*)}_k \btau_{0}\cdot 
{\bf T}^{\dagger}_{k} \nonumber \\
 &=& \sum_{\alpha=N,\Delta,\Delta^*} \sum_{\kappa={\rm s},0,\pm} \sum_{ab} t^
 {\eta\alpha}_{\kappa,ab}(\bfk_f, \bfk_i) \tau_{0\kappa}^{N} \tau_{k\kappa}^
 {\alpha \dag} \sigma_{0a}^{N} \sigma_{kb}^{\alpha \dag}.
\label{eq:tetak}
\end{eqnarray}
Note that $T_{k,-\kappa}=(T_{k}^{\dag})_{\kappa}^{\dag}=\tau^{(\Delta)\dag}
_{k\kappa}$ and so on from the definition (\ref{eq:tau}). The PWIA $t$-matrix
of Eq.~(\ref{eq:tpwopt}) now becomes
\begin{equation}
T_{nn_0}^{\rm PW}(\bfk_f, \bfk_i)=\sum_{\alpha} \sum_{\kappa} \sum_{ab} t_
{\kappa,ab}^{\eta\alpha}(\bfk_f, \bfk_i)F_{\kappa b}^{nn_0,\alpha}(\bfq) \tau_
{0\kappa}^{N}\sigma_{0a}^N,
\end{equation}
where the transition form factor of Eq.~(\ref{eq:trff}) is generalized as
\begin{equation}
F^{nn_0,\alpha}_{\kappa a}(\bfq)=\bra{\Psi_n}\sum_{k=1}^A \tau_{k\kappa}^ 
{\alpha \dag} \sigma_{ka}^{\alpha \dag}{\rm e}^{-i\bfq\cdot \bfr_{k}} 
\delta\left(\sum_{k'=1}^A\bfr_{k'}\right)\ket{\Psi_{n_0}}.
\end{equation}
Then the response function of Eq.~(\ref{eq:Rint}) becomes
\begin{equation}
R_{\lambda\kappa;ba}^{\beta\alpha}(q,\omega) = \frac{1}{2J_A+1}\sum_n 
{\sum_{n_0}}' F_{\kappa a}^{nn_0,\alpha}(\bfq)F_{\lambda b}^{nn_0,\beta *} 
(\bfq)\delta(\omega - ({\cal E}_{n} - {\cal E}_{n_0})),
\end{equation}
where only $R_{\kappa\kappa}$, $R_{0 \rm s}$ and $R_{\rm s0}$ remain finite 
due to the charge conservation.

Now the polarized cross sections $I_{\rm lab}D_i$ are expressed in PWIA as
\begin{eqnarray}
I_{\rm lab}^{\rm PW}D_i^{\rm PW}&=&K_{\rm lab}\sum_{\kappa\lambda}\bra{N'}
\tau_{0\kappa}^{N}\ket{N}\bra{N}\tau_ {0\lambda}^{N \dag}\ket{N'}\nonumber \\
&\times& \sum_{\alpha\beta}\sum_{ab}t_{\kappa,ia}^{\eta\alpha}(\bfk_f, \bfk_i)
R_{\lambda\kappa;ba}^{\beta\alpha}(q,\omega)t_{\lambda,ib}^{\eta\beta *} 
(\bfk_f, \bfk_i).
\label{eq:IDIIS}
\end{eqnarray}

The normalized response functions are usually introduced only for the diagonal
parts in the isospin space as\cite{ref:WA99}
\begin{equation}
\hat{R}_{\kappa;ba}^{\beta\alpha}(q,\omega)=\frac{1}{{\cal N}_{\kappa}}R_
{\kappa\kappa;ba}^{\beta\alpha}(q,\omega)
\end{equation}
with
\begin{equation}
{\cal N}_{-}=2N,\ \ {\cal N}_{0}={\cal N}_{\rm s}=A, \ \ {\cal N}_{+}=2Z,
\label{eq:calN}
\end{equation}
where $N$ and $Z$ are the numbers of neutrons and protons, respectively. The
normalization relation then becomes
\begin{equation}
\int \hat{R}_{\kappa;aa}^{NN}(q, \omega) d\omega \rightarrow 1 \ \ \mbox 
{for}\ \ q \rightarrow \infty,
\end{equation}
if the $\Delta$ component does not exist in the ground state.

To use Eq. (\ref{eq:IDIIS}), we need information about $t^{\eta\Delta}$ as 
well as $t^{\eta N}$, but to our knowledge it is not available for such a very
off-shell region. Suggested by the $\pi+\rho$ exchange model, we assume that
\begin{equation}
t^{\eta\Delta}_{\kappa, ii}=\frac{f_{\Delta}}{f_{N}}t^{\eta N}_{\kappa, ii}
\end{equation}
for $\kappa=0,+,-$ and $i=q, n, p$ where $f_N$ and $f_{\Delta}$ are the $\pi
NN$ and the $\pi N\Delta$ coupling constants, respectively. Then we define the
normalized isovector spin-longitudinal and -transverse response functions with
the $\Delta$ contribution, $\hat{R}_ {\kappa, \rm L}$ and $\hat{R}_{\kappa, 
\rm T}$, as\cite{ref:NI95}
\begin{eqnarray}
\hat{R}_{\kappa, \rm L}&=&\hat{R}^{NN}_{\kappa, qq}+2\left(\frac{f_{\Delta}}
{f_{N}}\right)\hat{R}^{\Delta N}_{\kappa, qq}+\left(\frac{f_{\Delta}}{f_{N}}
\right)^2\hat{R}^{\Delta\Delta}_{\kappa, qq}, \nonumber \\
\hat{R}_{\kappa, \rm T}&=&\hat{R}^{NN}_{\kappa, pp}+2\left(\frac{f_{\Delta}}
{f_{N}}\right)\hat{R}^{\Delta N}_{\kappa, pp}+\left(\frac{f_{\Delta}}{f_{N}}
\right)^2\hat{R}^{\Delta\Delta}_{\kappa, pp}
\label{eq:Rkappa}
\end{eqnarray}
in place of Eq.~(\ref{eq:Rhat}).

For the $(\vec{p}, \vec{n})$ reaction, Eq.~(\ref{eq:IDIwA}) are generalized as
\begin{eqnarray}
\tilde{I}_{\rm lab}\tilde{D}_q&=&4K_{\rm lab} N_{\rm eff} |E^{\eta}(\bfk_f, 
\bfk_i)|^2 \hat{R}_{-,\rm L}(q,\omega), \nonumber \\
\tilde{I}_{\rm lab}\tilde{D}_p&=&4K_{\rm lab} N_{\rm eff} |F^{\eta}(\bfk_f, 
\bfk_i)|^2 \hat{R}_{-,\rm T}(q,\omega)
\label{eq:IDipn}
\end{eqnarray}
by using Eqs.~(\ref{eq:IDIIS})--(\ref{eq:calN}), (\ref{eq:Rkappa}) and $\bra
{n} \tau_{0-}\ket{p}\bra{p}\tau_{0-}^{\dag}\ket{n}=2$. The effective neutron
number $N_{\rm eff}$ is defined in a similar way as that of Eq.~(\ref 
{eq:Aeff}). For the $(\vec{n}, \vec{p})$ reaction, $N_{\rm eff}$ should be 
replaced by the effective proton number $Z_{\rm eff}$ and $\hat{R}_ {-,\rm L
(T)}$ by $\hat{R} _{+,\rm L(T)}$. For the isovector part of the $(\vec{p}, 
\vec{p}')$ or $(\vec{n}, \vec{n}')$ scattering, $4N_ {\rm eff}$ should be 
replaced by $A_ {\rm eff}$ and $\hat {R}_{-,\rm L(T)}$ by $\hat {R}_{0,\rm L
(T)}$.

\section{DWIA formalism}

\label{sec:DWIA}

We now present our DWIA formalism. To avoid unnecessary complexity, we 
suppress the isospin and the $\Delta$ degrees of freedom until Sec.~\ref
{sec:DWID}

\subsection{DWIA $t$-matrix in momentum representation}

The matrix elements of the DWIA $t$-matrix in the $NA$ c.m.\ system are 
written as
\begin{equation}
\left[T_{nn_0}^{\rm DW}(\bfk_f, \bfk_i) \right]_{m_{s_f} m_{s_i}} =\bra{\chi_
{\bfk_f m_{s_f}}^{(-)}(\bfk') \Psi_n} \sum_{k=1}^A t_k(\bfk', \bfk) \ket{\Psi_
{n_0} \chi_{\bfk_i m_{s_i}}^{(+)}(\bfk)}
\end{equation}
in the momentum representation, where $\chi_{\bfk_i m_{s_i}}^{(+)}(\bfk)$
($\chi_{\bfk_f m_{s_f}}^{(-)}(\bfk')$) is the distorted wave in the momentum
space with the incident (outgoing) momentum $\bfk_i$ ($\bfk_f$) and the spin
projection $m_{s_i}$($m_{s_f}$) in the asymptotic region. An important 
difference from the previous section is that the momenta $\bfk$ and $\bfk'$ 
are now the integration variables, whereas in PWIA they are fixed as 
$\bfk=\bfk_i$ and $\bfk'=\bfk_f$. To clarify this difference we use the 
notations
\begin{equation}
\bfq^* = \bfk'-\bfk,
\end{equation}
\begin{equation}
\hat{\bfq}^* = \frac{\bfq^*}{\mid \bfq^* \mid}, \ \ \
\hat{\bfn}^* = \frac{\bfk \times \bfk'} {\mid \bfk \times \bfk' \mid}, \ \ \
\hat{\bfp}^* = \hat{\bfq}^* \times \hat{\bfn}^*
\end{equation}

As in the previous section, we again adopt the optimal factorization 
approximation and set $\bfq^* = \bfq_{\rm c}$, $R^{\rm \ell}_{\rm spin}=R^{\rm r}_
{\rm spin}=1$ and thus identify the $[\hat{\bfq}^*, \hat{\bfn}^*, \hat{\bfp}
^*]$ frame with the $[\hat{\bfq}_{\rm c}, \hat{\bfn}_{\rm c}, \hat{\bfp}_{\rm
c}]$ one. Then the $NN$ $t$-matrix in the optimal frame is written as
\begin{eqnarray}
t^{\eta}_k(\bfk',\bfk)&=& \sum_{a^*b^*}t^{\eta}_{a^*b^*}(\bfk',\bfk)\sigma_
{0a^*}\sigma_{kb^*} = J_{\eta}(\bfk',\bfk)\sum_{a^*b^*}t^{\rm cm}_{a^*b^*}
(\bkappa',\bkappa)\sigma_{0a^*}\sigma_{kb^*}\nonumber \\
   &=& J_{\eta}(\bfk',\bfk)\left[ A'\sigma_{00}\sigma_{k0} + B'\sigma_
   {0n^*}\sigma_{kn^*} + C'(\sigma_{0n^*}\sigma_{k0}+ \sigma_{00}\sigma_
   {kn^*}) \right.\nonumber \\
   &+& \left. E'\sigma_{0q^*}\sigma_{kq^*} + F'\sigma_{0p^*}\sigma_
   {kp^*}\right],
   \label{eq:qast}
\end{eqnarray}
where $a^*,b^*=0,q^*,n^*,p^*$ and the momenta $\bkappa$ and $\bkappa'$ are 
determined by $\bfk$ and $\bfk'$ through the coordinate transformation between
the optimal frame and the $NN$ c.m. frame.

\subsection{DWIA $t$-matrix in coordinate representation}

Since the distorted waves $\chi_{\bfk m_s}^{(\pm)}$ are usually calculated in
the coordinate space, we now move to the coordinate representation.

If $t^{\eta}(\bfk',\bfk)$ were a function of only $\bfq^*$, i.e. $t^{\eta}
(\bfk',\bfk)=t^{\eta}(\bfq^*)$, its coordinate representation would be local
and $T^{\rm DW}$ would be written as
\begin{equation}
\left[T_{nn_0}^{\rm DW}(\bfk_f, \bfk_i) \right]_{m_{s_f} m_{s_i}} = \bra{\chi_
{\bfk_f m_{s_f}}^{(-)}(\bfr) \Psi_n} \sum_{k=1}^A {\cal V}_k(\bfr-\bfr_k) \ket
{\Psi_{n_0} \chi_{\bfk_i m_{s_i}}^{(+)}(\bfr)}
\end{equation}
with
\begin{equation}
{\cal V}_k(\bfr-\bfr_k)=\int\frac{d^3\bfq^*}{(2\pi)^3}t^{\eta}_k(\bfq^*){\rm
e}^{i\bfq^*\cdot(\bfr-\bfr_k)}
\label{eq:Vk}
\end{equation}
and
\begin{equation}
\chi_{\bfk m_s}^{(\pm)}(\bfr)=\int\frac{d^3\bfk'}{(2\pi)^3}\chi_{\bfk m_s}^
{(\pm)}(\bfk'){\rm e}^{i\bfk'\cdot\bfr}.
\end{equation}
However, this is not the case in general. Neither the amplitudes $t^{\eta}_
{a^*b^*}(\bfk',\bfk)$ nor the directions $a^*,b^*=n^*,p^*$ in Eq.~(\ref
{eq:qast}) are not determined only by $\bfq^*$.

The amplitudes $t_{a'b'}^{\rm cm}(\bkappa', \bkappa)$ depend on both $\bfq_
{\rm c}=\bkappa'-\bkappa=\bfq^*$ and ${\bf Q}_{\rm c}=\bkappa'+\bkappa$ as 
well as the incident energy. Taking the same approximation as Love and 
Franey\cite{ref:LF81}, we suppress the ${\bf Q}_{\rm c}$ dependence and regard
the amplitudes as only functions of $\bfq^*$, namely $t_{a'b'}^{\rm cm}
(\bfq^*)$. We also replace the M{\"o}ller factor $J_{\eta}(\bfk', \bfk)$ by 
$\overline{J}_{\eta}=J_{\eta}(\bfk_f,\bfk_i)$ and thus it becomes independent
of $\bfk'$ and $\bfk$. This approximation works for small angle scattering at
high incident energy.

As for the spin parts, the relations $B'=F'$ and $C'=0$ were assumed in 
Ref.~\cite{ref:IKJG89}. Then they became only $\bfq^*$-dependent, because 
$\sigma_{0n^*}\sigma_{kn^*}+\sigma_{0p^*}\sigma_{kp^*}=(\bsigma_0\times\hat
{\bfq}^*)\cdot (\bsigma_k\times\hat{\bfq}^*)$. However, the approximation 
becomes poorer as the incident energy increases\cite{ref:BUG92,ref:AR94}. 
Therefore we rewrite and approximate $t^{\eta}_k(\bfk',\bfk)$ of Eq.~(\ref
{eq:qast}) as
\begin{eqnarray}
t^{\eta}_k(\bfk',\bfk) &\approx& t^{\eta}_k(\bfq^*)\nonumber \\
  &=& \overline{J}_{\eta}\left[ A'(q^*)\sigma_{00}\sigma_{k0} + E'(q^*)
  (\bsigma_{0}\cdot\hat{\bfq}^*)(\bsigma_{k}\cdot\hat{\bfq}^*) + F'(q^*)
  (\bsigma_{0}\times\hat{\bfq}^*)(\bsigma_{k}\times\hat{\bfq}^*) 
  \right.\nonumber \\
  &+& \left.(B'(q^*)-F'(q^*))\sigma_{0n}\sigma_{kn} + C'(q^*)(\sigma_{0n}
  \sigma_{k0} + \sigma_{00}\sigma_{kn})\right]
  \label{eq:tlocal}
\end{eqnarray}
where $\hat{\bfn}^*$ in the last two terms is replaced with the averaged 
normal vector $\hat{\bfn}$ of Eq.~(\ref{eq:direc-cm}), which is independent
of $\bfk$ and $\bfk'$.

We now obtain the local interaction ${\cal V}_k(\bfr-\bfr_k)$ in the 
coordinate space from Eq.~(\ref{eq:Vk}). Consequently the matrix elements of
the DWIA $t$-matrix can be written as
\begin{equation}
\left[T_{nn_0}^{\rm DW}(\bfk_f, \bfk_i)\right]_{m_{s_f} m_{s_i}} =\bra{\Psi_n}
{\cal S}_{m_{s_f} m_{s_i}} \ket{\Psi_{n_0}}
  \label{eq:sumSk}
\end{equation}
with
\begin{equation}
  {\cal S}_{m_{s_f} m_{s_i}} \equiv \sum_{k} \bra{\chi_{\bfk_f m_{s_f}}^{(-)}
  (\bfr)}{\cal V}_k(\bfr-\bfr_k) \ket{\chi_{\bfk_i m_{s_i}}^{(+)}(\bfr)}.
\label{eq:Sk}
\end{equation}

\subsection{Cross Sections and Spin Observables}
\label{sub:PT}

Inserting the DWIA $t$-matrix of Eq.~(\ref{eq:sumSk}) into Eqs.~(\ref
{eq:Icm2}) and (\ref{eq:SpOb}), we obtain the DWIA formulas for the cross 
sections and the spin observables. In this procedure we need to prepare the 
response functions for the operator ${\cal S}_{m_{s_f} m_{s_i}}$ of
Eq.~(\ref{eq:sumSk})
\begin{eqnarray}
\lefteqn{\left[R_{\cal S}(\omega)\right]_{m_{s_f}m_{s_i};m'_{s_f}m'_{s_i}} 
\equiv \frac{1}{2J_A+1}{\rm Tr'}\left[T_{m_{s_f}m_{s_i}}T^{\dagger}_{m'_{s_i}
m'_{s_f}}\right]}\nonumber \\
&=&\frac{1}{2J_A+1}\sum_n {\sum_{n_0}}' \left[T_{nn_0}\right]_{m_{s_f}m_{s_i}}
\left[T^{\dagger}_{n_0 n}\right]_{m'_{s_i}m'_{s_f}} \delta\left(\omega-({\cal
E}_n -{\cal E}_{n_0})\right) \nonumber \\
&=&\frac{1}{2J_A+1} {\sum_{n_0}}' \bra{\Psi_{n_0}} {\cal S}_{m'_{s_i} m'_
{s_f}}^{\dagger} \delta(\omega-(H_{A}-{\cal E}_{n_0})) {\cal S}_{m_{s_f} m_
{s_i}} \ket{\Psi_{n_0}}
\label{eq:RS}
\end{eqnarray}

Explicitly writing the operation Tr in Eqs.~(\ref{eq:Icm2}) and (\ref
{eq:SpOb}), we express the cross section and the spin observables by $R_{\cal
S}$ as
\begin{eqnarray}
I & = & \frac{K}{2}\left( \left[R_{\cal S}\right]_{++;++}+\left[R_{\cal S}
\right]_{+-;+-}+\left[R_{\cal S}\right]_{-+;-+}+\left[R_{\cal S}\right]_
{--;--} \right), \nonumber \\
I P_y & = & K{\rm Im}\left( \left[R_{\cal S}\right]_{++;-+}+\left[R_{\cal S}
\right]_{+-;--} \right), \nonumber \\
I A_y & = & K{\rm Im}\left( \left[R_{\cal S}\right]_{++;+-}+\left[R_{\cal S}
\right]_{-+;--} \right), \nonumber \\
I D_{xx} & = & K{\rm Re}\left( \left[R_{\cal S}\right]_{++;--}+\left[R_{\cal
S}\right]_{+-;-+} \right), \nonumber \\
I D_{xz} & = & K{\rm Re}\left( \left[R_{\cal S}\right]_{+-;++}-\left[R_{\cal
S}\right]_{--;-+} \right), \nonumber \\
I D_{yy} & = & K{\rm Re}\left( \left[R_{\cal S}\right]_{++;--}-\left[R_{\cal
S}\right]_{+-;-+} \right), \nonumber \\
I D_{zx} & = & K{\rm Re}\left( \left[R_{\cal S}\right]_{++;-+}-\left[R_{\cal
S}\right]_{--;+-} \right), \nonumber \\
I D_{zz} & = & \frac{K}{2}\left( \left[R_{\cal S}\right]_{++;++}-\left[R_
{\cal S}\right]_{+-;+-}-\left[R_{\cal S}\right]_{-+;-+}+\left[R_{\cal S}
\right]_{--;--} \right).
\label{eq:RS2ID}
\end{eqnarray}
where the suffices $+$ and $-$ means $m_{s}=\frac{1}{2}$ and $-\frac{1}{2}$,
respectively.

\section{Angular momentum representation}
\label{sec:ang}

Let us now move on to the angular momentum representation in which calculation
is actually carried out. We adopt the $[\hat{\bf x},\hat{\bf y},\hat{\bf z}]
$ frame and use the spherical tensor form $\sigma^s_{k\mu}$ for the spin 
operators and $\hat n_ {\mu}$ for the normal vector $\hat{\bfn}$, as follows:
\begin{equation}
\sigma^0_{k0} \equiv {\bf 1}_k\ \ \sigma^1_{k\ \pm 1}\equiv \mp \frac{\sigma_
{kx}\pm \ii \sigma_{ky}}{\sqrt{2}},\ \ \sigma^1_{k0}\equiv \sigma_{kz},
\end{equation}
\begin{equation}
\hat n_{\pm 1}=\mp\frac{\hat n_x \pm \ii \hat n_y}{\sqrt{2}} =-\frac{\ii} 
{\sqrt{2}},\ \ \hat n_0 = \hat n_z = 0.
\end{equation}

According to Eq. (\ref{eq:tlocal}), the interaction ${\cal V}_k(\bfr-\bfr_k)
$ can be decomposed as
\begin{eqnarray}
{\cal V}_k(\bfr-\bfr_k)&=&{\cal V}^A_k(\bfr-\bfr_k)+{\cal V}^E_k(\bfr-\bfr_k)
+{\cal V}^F_k(\bfr-\bfr_k) \nonumber \\
& &+{\cal V}^{B-F}_k(\bfr-\bfr_k)+{\cal V}^{C_1}_k(\bfr-\bfr_k)+{\cal V}^{C_2}
_k(\bfr-\bfr_k).
\label{eq:Vdr}
\end{eqnarray}
Each term of the r.h.s. is given in the angular momentum representation as

\noindent
$A$-term
\begin{eqnarray}
{\cal V}_k^A(\bfr-\bfr_k)
&=& \overline{J}_{\eta} \int \frac{d^3\bfq}{(2\pi)^3} A'(q) \sigma_{00} 
\sigma_{k0} {\rm e}^{\ii \bfq\cdot (\bfr-\bfr_k)} \nonumber \\
&=& \overline{J}_{\eta} \frac{2}{\pi}\sum_{lm}\int_0^{\infty}j_l(qr_k)A'(q)j_l
(qr) q^2 dq {\left[ \ii^l Y_l(\hat{\bfr}_k)\times \sigma_k^0 \right]^l_m}^
{\dagger} \left[ \ii^l Y_l(\hat{\bfr})\times  \sigma_0^0 \right]^l_m \nonumber
\\
&\equiv& \sum_{lm} V^A_l(r_k, r) {\left[ \ii^l Y_l(\hat{\bfr}_k)\times 
\sigma_k^0 \right]^l_m}^{\dagger} \left[ \ii^l Y_l(\hat{\bfr})\times  
\sigma_0^0 \right]^l_m,
\label{eq:Va}
\end{eqnarray}

\noindent
$E$-term
\begin{eqnarray}
{\cal V}_k^E(\bfr-\bfr_k)
&=& \overline{J}_{\eta} \int \frac{d^3\bfq}{(2\pi)^3} E'(q) 
(\bsigma_0\cdot\hat\bfq)(\bsigma_k\cdot\hat\bfq) {\rm e}^{\ii \bfq\cdot 
(\bfr-\bfr_k)} \nonumber \\
&=& \overline{J}_{\eta} \frac{2}{\pi}\sum_{JM}\sum_{ll'} a_{Jl} a_{Jl'} 
\int_0^{\infty} j_l(qr_k)E'(q) j_{l'}(qr) q^2 dq \nonumber \\
  & &\times{\left[ \ii^l Y_l(\hat{\bfr}_k)\times\sigma_k^1\right]^J_M} ^
  {\dagger}\left[ \ii^{l'} Y_{l'}(\hat{\bfr})\times\sigma_0^1\right]^J_M 
  \nonumber \\
&\equiv& \sum_{JM}\sum_{ll'} V^E_{Jll'}(r_k, r) {\left[ \ii^l Y_l(\hat{\bfr}
_k)\times\sigma_k^1\right]^J_M}^{\dagger} \left[ \ii^{l'} Y_{l'}(\hat{\bfr})
\times\sigma_0^1\right]^J_M,
\end{eqnarray}

\noindent
$F$-term
\begin{eqnarray}
{\cal V}_k^F(\bfr-\bfr_k)
&=& \overline{J}_{\eta} \int \frac{d^3\bfq}{(2\pi)^3} F'(q) 
(\bsigma_0\times\bfq) (\bsigma_k\times\bfq) {\rm e}^{\ii \bfq\cdot (\bfr-\bfr_k)}
\nonumber \\
&=& \overline{J}_{\eta} \frac{2}{\pi}\sum_{JM}\sum_{ll'} (\delta_{ll'}-a_{Jl}
a_{Jl'}) \int_0^{\infty} j_l(qr_k)F'(q) j_{l'}(qr) q^2 dq \nonumber \\
  & &\times{\left[ \ii^l Y_l(\hat{\bfr}_k)\times\sigma_k^1\right]^J_M} ^
  {\dagger}\left[ \ii^{l'} Y_{l'}(\hat{\bfr})\times\sigma_0^1\right]^J_M 
  \nonumber \\
&\equiv& \sum_{JM}\sum_{ll'} V^F_{Jll'}(r_k, r) {\left[ \ii^l Y_l(\hat{\bfr}
_k)\times\sigma_k^1\right]^J_M}^{\dagger} \left[ \ii^{l'} Y_{l'}(\hat{\bfr})
\times\sigma_0^1\right]^J_M,
\label{eq:FTVF}
\end{eqnarray}

\noindent
$(B-F)$-term
\begin{eqnarray}
{\cal V}_k^{B-F}(\bfr-\bfr_k)
&=& \overline{J}_{\eta} \int \frac{d^3\bfq}{(2\pi)^3} \left(B'(q)-F'(q)\right)
\sigma_{0n} \sigma_{kn} {\rm e}^{\ii \bfq\cdot (\bfr-\bfr_k)} \nonumber \\
&=& \overline{J}_{\eta} \frac{2}{\pi}\sum_{lm}\sum_{JM}\sum_{J'M'}\sum_
{\mu\mu'} \langle lm\ 1\mu | JM\rangle \langle lm\ 1\mu' | J'M'\rangle \hat{n}
_{\mu} \hat{n}_{\mu'}^{\dagger}\nonumber \\
  & &\times\int_0^{\infty}j_l(qr_k)\left(B'(q)-F'(q)\right) j_l(qr) {q}^2 dq
  \nonumber \\
  & &\times{\left[ \ii^l Y_l(\hat{\bfr}_k)\times\sigma_k^1\right]^J_M} ^
  {\dagger}\left[ \ii^l Y_l(\hat{\bfr})\times\sigma_0^1\right]^{J'}_{M'}
\nonumber \\
&\equiv& \sum_{JM}\sum_{J'M'}\sum_{l} V^{B-F}_{lJMJ'M'}(r_k, r) {\left[ \ii^l
Y_l(\hat{\bfr}_k)\times\sigma_k^1\right]^J_M} ^{\dagger}\left[ \ii^l Y_l(\hat
{\bfr})\times\sigma_0^1\right]^{J'}_{M'}
\end{eqnarray}

\noindent
$C_1$-term
\begin{eqnarray}
{\cal V}_k^{C_1}(\bfr-\bfr_k)
&=& \overline{J}_{\eta} \int \frac{d^3\bfq}{(2\pi)^3} C'(q) \sigma_{0n} 
\sigma_{k0} {\rm e}^{\ii \bfq\cdot (\bfr-\bfr_k)} \nonumber \\
&=& \overline{J}_{\eta} \frac{2}{\pi}\sum_{lm}\sum_{JM} (-)^{M-m}\hat{n}_{-
(M-m)} \langle lm\ 1\ M-m | JM\rangle \nonumber \\
  & &\times\int_0^{\infty}j_l(qr_k)C'(q) j_l(qr) q^2 dq \nonumber \\
  & &\times{\left[ \ii^l Y_l(\hat{\bfr}_k)\times\sigma_k^0\right]^l_m} ^
  {\dagger}\left[ \ii^l Y_l(\hat{\bfr})\times\sigma_0^1\right]^J_M
\nonumber \\
&\equiv& -\sum_{lm}\sum_{JM} V^C_{lmJM}(r_k, r) {\left[ \ii^l Y_l(\hat{\bfr}
_k)\times\sigma_k^0\right]^l_m} ^{\dagger}\left[ \ii^l Y_l(\hat{\bfr})
\times\sigma_0^1\right]^J_M,
\end{eqnarray}

\noindent
$C_2$-term
\begin{eqnarray}
{\cal V}_k^{C_2}(\bfr-\bfr_k)
&=& \overline{J}_{\eta} \int \frac{d^3\bfq}{(2\pi)^3} C'(q) \sigma_{00}\sigma_
{kn}{\rm e}^{\ii \bfq\cdot (\bfr-\bfr_k)} \nonumber \\
&=& \overline{J}_{\eta} \frac{2}{\pi}\sum_{lm}\sum_{JM}\hat{n}_{(M-m)} \langle
lm\ 1\ M-m | JM\rangle \nonumber \\
  & &\times\int_0^{\infty}j_l(qr_k)C'(q) j_l(qr) {q}^2 dq \nonumber \\
  & &\times{\left[ \ii^l Y_l(\hat{\bfr}_k)\times\sigma_k^1\right]^J_M} ^
  {\dagger}\left[ \ii^l Y_l(\hat{\bfr})\times\sigma_0^0\right]^l_m
\nonumber \\
&=& \sum_{lm}\sum_{JM} V^C_{lmJM}(r_k, r) {\left[ \ii^l Y_l(\hat{\bfr}_k)
\times\sigma_k^1\right]^J_M} ^{\dagger}\left[ \ii^l Y_l(\hat{\bfr})
\times\sigma_0^0\right]^l_m,
  \label{eq:Vc2}
\end{eqnarray}
where $j_l(x)$ is the spherical Bessel function of the order $l$, $\langle l
m s \mu | J M \rangle$ is the Clebsch-Gordan coefficient and $a_{Jl}\equiv 
\langle J\ 0\ 1\ 0 | l\ 0\rangle.$ The spherical tensor product is expressed
as $\left[ \ii^l Y_l\times\sigma^s\right]^J_M\equiv\sum_{m\mu} \langle l m s
\mu | J M \rangle \ii^l Y_{lm}(\Omega) \sigma^s_{\mu}$.

For an optical potential with a spin-orbit force, the distorted wave $\chi^
{(+)}_{{\bf k}_i m_{s_i}} ({\bf r})$ is expressed in the angular momentum 
representation as\cite{ref:Sat83}
\begin{eqnarray}
\chi^{(+)}_{\bfk_i m_{s_i}}(\bfr) & = & \sum_{m_{s_i}'}\chi^{(+)}_{m_{s_i}' 
m_{s_i}} ({\bf k}_i,{\bf r}) \ket{m_{s_i}'} \nonumber \\
& = & {4\pi \over k_ir} \sum_{l_i m_i m_i' j_i} \sum_{m_{s_i}' m_{s_i}} 
\langle l_i m_i s m_{s_i} | j_i m_{j_i}\rangle Y^{\ast}_{l_i m_i}(\hat {\bf 
k}_i) \nonumber \\
& & \times \langle l_i m_i' s m_{s_i}' | j_i m_{j_i}\rangle Y_{l_i m_i'}(\hat
{\bf r})i^{l_i}e^{\ii\sigma_{l_i}} u^{(+)}_{l_i j_i}(k_i,r) \ket{m_{s_i}'},
\label{eq:chi}
\end{eqnarray}
where $\ket{m_s}$ is the intrinsic spin state with the $z$-projection $m_s$ 
and $\sigma_l$ is the Coulomb phase shift. The radial part of the distorted 
wave $u^{(+)}_{lj}(k,r)$ has the asymptotic behavior
\begin{equation}
u^{(+)}_{lj}(k,r) \sim e^{i\delta_{lj}}\sin(kr-\eta_{\rm C} \ln2kr-{l\pi \over
2}+\sigma_l + \delta_{lj}),
\end{equation}
where $\eta_{\rm C}$ is the Sommerfeld parameter and $\delta_{lj}$ is the 
nuclear phase shift.

Using Eqs.~(\ref{eq:Vdr}) and (\ref{eq:chi}) together with Eqs.~(\ref{eq:Va})
-- (\ref{eq:Vc2}), we can write as
\begin{eqnarray}
\bra{\chi^{(-)}_{\bfk_f m_{s_f}}(\bfr)}{\cal V}_k(\bfr-\bfr_k)\ket{\chi^{(+)}
_{\bfk_i m_{s_i}}(\bfr)}
&=& \sum_{lsJM} {\cal S}^{m_{s_f}m_{s_i}}_{(ls)JM}(\bfk_f, \bfk_i; r_k) {\left
[ \ii^l Y_l(\hat{\bfr}_k)\times\sigma_k^s\right]^J_M}^{\dagger},
   \label{eq:Samp}
\end{eqnarray}
where
\begin{eqnarray}
{\cal S}^{m_{s_f}m_{s_i}}_{(J0)JM}(\bfk_f, \bfk_i; r_k)
&=& \int_0^{\infty} r^2dr V_J^A(r_k,r) f_{(J0)JM}^{m_{s_f}m_{s_i}}(\bfk_f, 
\bfk_i; r) \nonumber \\
&& - \sum_{lm} \int_0^{\infty} r^2dr V_{lmJM}^C(r_k,r) f_{(l1)JM}^{m_{s_f}m_
{s_i}}(\bfk_f, \bfk_i; r)
   \label{eq:sJ0JM}
   \\
{\cal S}^{m_{s_f}m_{s_i}}_{(l1)JM}(\bfk_f, \bfk_i; r_k)
&=& \sum_{l'} \int_0^{\infty} r^2dr \left\{ V_{Jll'}^{E}(r_k,r) + V_{Jll'}^{F}
(r_k,r)\right\}f_{(l'1)JM}^{m_{s_f} m_{s_i}}(\bfk_f, \bfk_i; r) \nonumber \\
&& + \sum_{J'M'} \int_0^{\infty} r^2dr V_{lJMJ'M'}^{B-F}(r_k,r) f_{(l1)J'M'}
^{m_{s_f}m_{s_i}}(\bfk_f, \bfk_i; r) \nonumber \\
&& + \sum_{m} \int_0^{\infty} r^2dr V_{lmJM}^{C}(r_k,r) f_{(l0)lm}^{m_{s_f}m_
{s_i}}(\bfk_f, \bfk_i; r) \label{eq:sl1JM}
\end{eqnarray}
with
\begin{eqnarray}
\lefteqn{f^{m_{s_f} m_{s_i}}_{(ls)JM}({\bf k}_i,{\bf k}_f;r)}\\
 & \equiv & \sum_{m_{s_f}' m_{s_i}'} \int d\Omega_{{\bf r}} \chi^{(-)\ast}_{m_
 {s_f}' m_{s_f}} ({\bf k}_f,{\bf r})\bra{m_{s_f}'} \left[ i^l Y_l(\hat {\bf 
 r}) \times \sigma_0^s \right]^J_M \ket{m_{s_i}'} \chi^{(+)}_{m_{s_i}' m_
 {s_i}} ({\bf k}_i,{\bf r}) \nonumber \\
& = & {\sqrt{24\pi} \over k_i k_f r^2}\sqrt{(2J+1)(2l+1)} \sum_{l_i j_i} \sum_
{l_f j_f}i^{l_i+l-l_f}e^{i (\sigma_{l_f}+\sigma_{l_i})} \nonumber \\
& & \times u^{(+)}_{l_f j_f}(k_f,r) u^{(+)}_{l_i j_i}(k_i,r) (2l_i+1)\sqrt
{(2j_i+1)(2l_f+1)}\langle l_i 0 l 0 | l_f 0\rangle \nonumber \\
& & \times \left\{ \begin{array}{ccc} l_f & s_f & j_f \\ l_i & s_i & j_i \\ 
l & s & J \end{array} \right\} \langle l_i 0 s_i m_{s_i} | j_i m_{s_i}\rangle
\langle l_f m_f s_f m_{s_f} | j_f m_{s_i}+M\rangle \nonumber \\
& & \times \langle j_i m_{s_i} J M | j_f m_{s_i}+M\rangle (-)^{m_f+| m_f | 
\over 2} \sqrt{(l_f-| m_f |)! \over (l_f+| m_f |)!}P^{| m_f |}_{l_f} (\cos 
\theta).
\end{eqnarray}

We can now write ${\cal S}_{m_{s_f} m_{s_i}}$ of Eq.~(\ref{eq:Sk}) as
\begin{equation}
{\cal S}_{m_{s_f} m_{s_i}}=\sum_{lsJM} \int_0^{\infty} {\cal S}^{m_{s_f} m_
{s_i}}_{(ls)JM}({\bf k}_i,{\bf k}_f;r) \rho_{(ls)JM}^{\dagger}(r) r^2dr,
\label{eq:sumSk2}
\end{equation}
where
\begin{equation}
\rho_{(ls)JM}(r) = \sum^A_{k=1} \frac{\delta (r-r_k)}{r r_k} \left[ \ii^l Y_l
(\hat {\bf r}_k) \times \sigma^s_k \right]^J_M \delta\left(\sum_{k'=1}^A \bfr_
{k'}\right),
\label{eq:rhor}
\end{equation}
which is the radial part of the transition density
\begin{eqnarray}
\rho^s_{\mu} ({\bf r}) & \equiv & \sum^A_{k=1}\sigma^s_{k\mu} \delta ({\bf r}
-{\bf r}_k) \delta\left(\sum_{k'=1}^A \bfr_{k'}\right)\nonumber \\
& = & \sum_{lsJM} \langle lm s\mu | JM\rangle \rho_{(ls)JM}(r) \left[ \ii^l Y_{lm} 
(\hat {\bf r}) \right]^{\ast}.
\label{eq:rho}
\end{eqnarray}

\section{Calculation of Response Functions}
\label{sec:crpa}

In Sec.~\ref{sec:DWIA}, we presented a method of calculating the cross 
sections and the spin observables by using DWIA. It was shown that what we 
need are the response functions $R_{\cal S}(\omega)$ of Eq.~(\ref{eq:RS}). 
Here we describe a way of calculating them.

\subsection{Polarization propagator}

Following Refs.~\cite{ref:ADM85} and \cite{ref:NI95}, we introduce the 
polarization propagators for the spin-dependent transition density in the 
angular momentum representation as
\begin{eqnarray}
\Pi_{(l's')J'M';(ls)JM} (r',r;\omega) & \equiv & \frac{1}{2J_A+1}\sum_n {\sum_
{n_0}}' \left[ {\bra{\Psi_{n_0}} \rho_{(l's')J'M'} (r') \ket{\Psi_{n}} \bra
{\Psi_{n}} \rho^{\dagger}_{(ls)JM} (r) \ket{\Psi_{n_0}} \over \omega-({\cal 
E}_{n}-{\cal E}_{n_0})+\ii\eta}\right. \nonumber \\
& & \left. +{\bra{\Psi_{n_0}} \rho^{\dagger}_{(ls)JM}(r) \ket{\Psi_{n}} \bra
{\Psi_{n}} \rho_{(l's')J'M'} (r') \ket{\Psi_{n_0}} \over -\omega-({\cal E}_{n}
-{\cal E}_{n_0})+\ii\eta} \right] \nonumber \\
& = & \frac{1}{2J_A+1}{\sum_{n_0}}'\bra{\Psi_{n_0}} \left[ \rho_{(l's')J'M'}
(r') {1 \over \omega-(H_{A}-{\cal E}_{n_0})+\ii\eta} \rho^{\dagger}_{(ls)JM}
(r) \right. \nonumber \\
& & \left. + \rho^{\dagger}_{(ls)JM}(r){1 \over -\omega-(H_{A}-{\cal E}_{n_0})
+\ii\eta} \rho_{(l's')J'M'}(r') \right] \ket{\Psi_{n_0}}\nonumber \\
& = & \delta_{JJ'}\delta_{MM'}\Pi_{J(l's')(ls)} (r',r;\omega).
\end{eqnarray}
By using the Wigner-Eckart theorem, we can prove that $\Pi_{(l's')J'M';(ls)JM}
$ are diagonal with respect to $J$ and $M$ and independent of $M$. The 
corresponding response functions are defined as
\begin{eqnarray}
R_{J(l's')(ls)} (r',r;\omega) &\equiv& \frac{1}{2J_A+1}\sum_n {\sum_{n_0}}' 
\bra{\Psi_{n_0}} \rho_{(l's')JM} (r') \ket{\Psi_{n}} \bra{\Psi_{n}} \rho^
{\dagger}_{(ls)JM}(r) \ket{\Psi_{n_0}} \delta\left(\omega-({\cal E}_{n}-{\cal
E}_{n_0})\right) \nonumber \\
&=& -\frac{1}{\pi}{\rm Im} \Pi_{J(l's')(ls)}(r',r;\omega).
\label{eq:RJlsls}
\end{eqnarray}

Using Eqs.~(\ref{eq:sumSk2}) and (\ref{eq:rhor}), we obtain the response 
function $R_{\cal S}(\omega)$ of Eq.~(\ref{eq:RS}) as
\begin{eqnarray}
\left[R_{\cal S}(\omega)\right]_{m_{s_f} m_{s_i};m'_{s_f} m'_{s_i}}
&=& \sum_{JM} \sum_{ls} \sum_{l's'} \int_0^{\infty} r^2 dr \int_0^{\infty} 
r'^2 dr' {\cal S}^{m'_{s_f} m'_{s_i} \ast}_{(l's')JM}(\bfk_f, \bfk_i;r') 
\nonumber \\
&& \times R_{J(l's')(ls)}(r',r;\omega) {\cal S}^{m_{s_f} m_{s_i}}_{(ls)JM}
(\bfk_f, \bfk_i;r).
\label{eq:Rs}
\end{eqnarray}
Inserting these $R_{\cal S}$ into Eq.~(\ref{eq:RS2ID}), we can calculate the
double differential cross section and the spin obsevables.

In the following we consider only the case where the target nucleus is a 
doubly closed shell. Therefore we can set $J_A=0$ and omit the summation over
$n_0$, and denote ${\cal E}_{n_0}$ and $\Psi_{n_0}$ simply by ${\cal E}_0$ and
$\Psi_0$, respectively.

\subsection{Single particle model}

A simple approximation to evaluate the polarization propagators is the 
single-particle model. In this model $H_{A}$ is replaced by the 
single-particle-model Hamiltonian
\begin{equation}
H^{(0)}=\sum_k \left(-\frac{\nabla^2_k}{2m}+U_k\right)+\Delta E_0
\end{equation}
where $m=[(A-1)/A] m_N$, $U_k$ is the mean field for the $k$-th nucleon, and
$\Delta E_0$ consists of the masses of the constituent nucleons and the energy
shift to give the correct total Hartree-Fock energy of the target ground 
state. Since there appears only the energy difference between the target 
ground state and the final state of the residual nucleus for the response 
functions, we discard the term $\Delta E_0$ from now on.

The mean field $U_k$ is determined as the Hartree-Fock field for the target 
nucleus and the single particle energy $\epsilon_{\alpha}$ is given by the 
equation
\begin{equation}
\left(-\frac{\nabla^2}{2m}+U\right)\phi_{\alpha}=\epsilon_{\alpha}\phi_{\alpha}.
\label{eq:sph}
\end{equation}
We write $\epsilon_{\alpha}$ of the unoccupied state by $\epsilon_p$ and that
of the occupied state by $\epsilon_h$. Note that $\epsilon_{\alpha}$ is 
measured from one-nucleon separation threshold of the target.

\subsubsection{Polarization propagator}

The polarization propagator is given by
\begin{eqnarray}
\Pi^{(0)}_{J(l's')(ls)}(r',r;\omega)&=&\sum_{ph}\left[\frac{\bra{\Phi_0}\rho_
{(l's')JM}(r')\ket{ph}\bra{ph}\rho^{\dagger}_{(ls)JM}(r)\ket{\Phi_0}}{\omega-
(\epsilon_{p}-\epsilon_{h})+\ii\eta}\right.\nonumber \\
& &\left. +\frac{\bra{\Phi_0}\rho^{\dagger}_{(ls)JM}(r)\ket{ph}\bra{ph}\rho_
{(l's')JM}(r)\ket{\Phi_0}}{-\omega-(\epsilon_{p}-\epsilon_{h})+\ii\eta}\right]
,
\end{eqnarray}
where $\ket{ph}$ denotes a one-particle-one-hole state and $\Phi_0$ is the 
ground state of the target in the single particle model. This is called the 
{\it uncorrelated} polarization propagator.

The infinite sum over $p$ can be handled by the single-particle Green's 
function\cite{ref:SB75} as
\begin{eqnarray}
\Pi^{(0)}_{J(l's')(ls)}(r',r;\omega) &=&  \sum_{ph} \left[{\cal B}_{(l's')J}
(h,p) \frac{u^*_h(r')}{r'} \frac{g_p(r',r;\omega+\epsilon_h)}{r'r} \frac{u_h
(r)}{r} {\cal B}^*_{(ls)J}(h,p) \right.\nonumber \\
& & \left. +{\cal B}_{(l's')J}(p,h) \frac{u^*_h(r)}{r} \frac{g_p
(r,r';-\omega+\epsilon_h)}{rr'} \frac{u_h(r')}{r'} {\cal B}^*_{(ls)J}(p,h)
\right]
\label{eq:Pi0}
\end{eqnarray}
where $p$ and $h$ denote the sets of the single-particle quantum numbers $p=
(l_p s_p j_p)$ and $h=(n_h l_h s_h j_h)$, respectively, and
\begin{equation}
{\cal B}_{(ls)J}(a,b)=\sqrt{2j_a+1}\sqrt{2j_b+1}\left\{
\begin{array}{ccc}
l_a & s_a & j_a \\ l_b & s_b & j_b \\ l & s & J
\end{array}
\right\} \langle l_a || \ii^l Y_l || l_b \rangle\langle s_a || \sigma^s || s_b
\rangle .
\end{equation}
The radial part of the single-particle Green's function is given by
\begin{equation}
g_p(r',r;E)=\frac{2m}{W(f_p,h_p)}f_p(r_{<},E)h_p(r_{>},E)
\end{equation}
where $W$ is the Wronskian and $r_{<(>)}$ denotes the smaller (larger) one of
$r$ and $r'$. The radial wave functions $u_h$, $f_p$ and $h_p$ are the bound
state, the regular and the irregular solution of Eq.~(\ref{eq:sph}), 
respectively.

\subsubsection{Spreading widths}

To take account of the spreading width of the particle states, we adopt a 
complex potential for the single-particle potential $U$ as in Refs.\cite
{ref:IKJG89},
\begin{equation}
U(r)=V(r)+\ii W(r),
\end{equation}
while we use the real potential for the hole states. For such a choice, the 
orthogonality between the hole and the particle wave functions is destroyed.
We therefore utilize the orthogonality condition prescription\cite{ref:Izu83}
, details of which are explained in Ref.\cite{ref:KI91}.

To take account of the spreading width $\gamma_h$ of the hole states, we replace
the hole energy $\epsilon_h$ by the complex energy\cite{ref:ITA96}
\begin{equation}
\tilde{\epsilon}_h=\epsilon_h+\ii\frac{\gamma_h}{2}.
\label{eq:epsh}
\end{equation}

\subsubsection{Effective mass}
\label{sec:kmass}

In principle the single-particle potential $U$ is non-local, and thus the 
Schr\"odinger equation for the single-particle states is
\begin{equation}
-\frac{1}{2m}\nabla^2\phi(\bfr)+\int d\bfr'U(\bfr,\bfr')\phi(\bfr')=\epsilon
\phi(\bfr).
\end{equation}
We deal with this non-locality by an effective mass approximation\cite
{ref:PB62,ref:Sat83}. We first introduce the Wigner transformation of 
the potential $U(\bfr,\bfr')$ as
\begin{equation}
\tilde{U}(\bfr,\bfq)=\int d{\bf s} U(\bfr + \frac{\bf s}{2}, \bfr - \frac{\bf
s}{2}){\rm e}^{\-\ii \bfq\cdot{\bf s}}.
\end{equation}
We assume that $\tilde{U}$ is spherically symmetric with respect to $\bfr$ and
$\bfq$, respectively,  and write it as $\tilde{U}(r, q^2)$. We expand it with
respect to $q^2$ around the square of the local momentum $\kappa(r)$
\begin{equation}
\kappa^2(r)=2m\epsilon-\tilde{U}(r, \kappa^2(r)),
\end{equation}
and we keep only the terms up to the first order. Introducing the 
radial-dependent effective mass $m^*(r)$ as
\begin{equation}
\frac{m^*(r)}{m}=f(r)=\left(1+2m\frac{\partial \tilde{U}(r, \kappa^2(r))}
{\partial\kappa^2}\right)^{-1},
\end{equation}
we obtain the equation
\begin{equation}
\left[\nabla\left(-\frac{1}{2m^*(r)}\nabla\right)+U_{\rm L}(r)-\frac{1}{8}
\left(\nabla^2\frac{1}{m^*(r)}\right)\right] \phi(\bfr)=\epsilon\phi(\bfr),
\end{equation}
where the local potential $U_{\rm L}(r)$ is
\begin{equation}
U_{\rm L}(r)=\frac{1}{f(r)}(\tilde{U}(r,\kappa^2(r))-\epsilon)+\epsilon. \end
{equation} Since we do not know about $U(\bfr,\bfr')$, we determine $U_{\rm 
L}(r)$ and $f(r)$ phenomenologically. Note that here we only consider the 
$k$-mass but not the $E$-mass.

\subsection{Ring approximation}

Next let us take account of the nuclear correlation by means of RPA. To treat
the exchange terms of RPA we further utilize the ring approximation, which 
replaces the exchange effects by a contact interaction. The polarization 
propagator then satisfies the RPA equation\cite{ref:FW71,ref:ADM85,ref:NI95}
\begin{eqnarray}
\lefteqn{\Pi_{J(l's')(ls)}(r',r;\omega)=\Pi^{(0)}_{J(l's')(ls)}(r',r;\omega)}
\nonumber \\
& &+\sum_{l_1 s_1 l_2 s_2}\int_0^{\infty} r_1^2 dr_1 r_2^2 dr_2 \Pi^{(0)}_{J
(l's')(l_1 s_1)}(r',r_1;\omega) W_{J (l_1 s_1) (l_2 s_2)}(r_1, r_2) \Pi_{J(l_2
s_2)(ls)}(r_2,r;\omega)
\end{eqnarray}
where $W_{J (l_1s_1) (l_2s_2)}(r_1, r_2)$ is the radial part of the effective
interaction as
\begin{equation}
V(\bfr_1,\bfr_2)=\sum_{JM} \sum_{l_1s_1l_2s_2} \left[ \ii^{l_1} Y_{l_1}(\hat
{\bf r}_1) \times \sigma^{s_1}_1 \right]^{J\dagger}_M W_{J (l_1s_1) (l_2s_2)}
(r_1, r_2) \left[ \ii^{l_2} Y_{l_2}(\hat {\bf r}_2) \times \sigma^{s_2}_2 
\right]^J_M.
\end{equation}

\subsection{Isospin and $\Delta$ degrees of freedom}
\label{sec:DWID}

From Sec.~\ref{sec:DWIA} up to here we have suppressed the isospin and the 
$\Delta$ degrees of freedom, but of course we have to include them in the 
actual calculation. Here we summarize the final formulas including these 
degrees of freedom. As was discussed in Sec.~\ref{sec:isodel}, the relevant 
quantities carry the superscript $\alpha$ ($=N, \Delta, \Delta^*)$ and the
isospin subscript $\kappa$. The extension is straightforward and easily
understood.

The transition density operator (\ref{eq:rhor}) is to be
\begin{equation}
\rho^{\alpha}_{\kappa(ls)JM}(r) = \sum^A_{k=1} \frac{\delta (r-r_k)}{r r_k} 
\tau^{\alpha}_{k\kappa} \left[ \ii^l Y_l(\hat {\bf r}_k) \times \sigma^{\alpha
s}_k \right]^J_M \delta\left(\sum_{k'=1}^A \bfr_{k'}\right),
\end{equation}
and thus the response function (\ref{eq:RJlsls}) becomes
\begin{eqnarray}
R^{\beta\alpha}_{\lambda\kappa,J(l's')(ls)} (r',r;\omega) &\equiv& \frac{1}
{2J_A+1}\sum_n {\sum_{n_0}}' \bra{\Psi_{n_0}} \rho^{\beta}_{\lambda (l's')JM}
(r') \ket{\Psi_{n}} \bra{\Psi_{n}} \rho^{\alpha \dagger}_{\kappa (ls)JM}(r) 
\ket{\Psi_{n_0}} \nonumber \\
& &\times\delta\left(\omega-({\cal E}_{n}-{\cal E}_{n_0})\right).
\end{eqnarray}
Further details are given in Ref.~\cite{ref:NI95}.

The $NN$ $t$-matrix was generalized as Eq.~(\ref{eq:tetak}). Correspondingly,
the source function (\ref{eq:Samp}) is generalized as
\begin{eqnarray}
\lefteqn{ \bra{\chi_{\bfk_f m_{s_f}}^{N' (-)}(\bfr)} {\cal V}_k(\bfr-\bfr_k)
\ket{\chi_{\bfk_i m_{s_i}}^{N (+)}(\bfr)} } \nonumber \\
&=& \sum_{\kappa} \bra{N'} \tau^{N}_{0\kappa} \ket{N} \sum_{\alpha} \sum_
{lsJM} {\cal S}^{\alpha m_{s_f}m_{s_i}}_{\kappa (ls)JM}(N'\bfk_f, N\bfk_i; 
r_k) {\tau^{\alpha}_{k\kappa}}^{\dagger}{\left[ \ii^l Y_l(\hat{\bfr}_k)
\times\sigma_k^{\alpha s}\right]^J_M}^{\dagger}
\end{eqnarray}
with $N, N'=p$ or $n$, which specifies the reaction type. Here the 
generalized forms of Eqs.~(\ref{eq:sJ0JM}) and (\ref{eq:sl1JM}) are
\begin{equation}
{\cal S}^{\alpha m_{s_f}m_{s_i}}_{\kappa (ls)JM}(N'\bfk_f, N\bfk_i; r_k) = 
\sum \int_0^{\infty} r^2 dr V^{\alpha X}_{\kappa \cdots} (r_k, r) f^{m_{s_f}
m_{s_i}}_{\cdots} (N'\bfk_f, N\bfk_i; r)
\end{equation}
with
\begin{eqnarray}
\lefteqn{f^{m_{s_f} m_{s_i}}_{(ls)JM}(N'{\bf k}_f, N{\bf k}_i;r)}\nonumber\\
 & = & \sum_{m_{s_f}' m_{s_i}'} \int d\Omega_{{\bf r}} \chi^{N' (-)\ast}_{m_
{s_f}' m_{s_f}} ({\bf k}_f,{\bf r})\bra{m_{s_f}'} \left[ i^l Y_l(\hat {\bf r})
\times \sigma_0^{Ns} \right]^J_M \ket{m_{s_i}'} \chi^{N (+)}_{m_{s_i}' m_{s_i}}
({\bf k}_i,{\bf r}),
\end{eqnarray}
where $V^{\alpha X}_{\kappa \cdots}$ are the generalized forms of $V^A_J, V^C_
{lmJM}, V^E_{Jll'}, V^F_{Jll'}$ and $V^{B-F}_{lJMJ'M'}$.

Finally, we obtain
\begin{eqnarray}
\lefteqn{\left[R^{(N,N')}_{\cal S}(\omega)\right]_{m_{s_f} m_{s_i};m'_{s_f} 
m'_{s_i}}=} \nonumber \\
& &\ \ \ \sum_{\lambda\kappa} \bra{N} \tau^{N \dag}_{0\lambda} \ket{N'}\bra
{N'} \tau^{N}_{0\kappa} \ket{N} \sum_{\alpha\beta} \sum_{JM} \sum_{ls} \sum_
{l's'} \int_0^{\infty} r^2 dr \int_0^{\infty} r'^2 dr' \nonumber \\
& &\ \ \ {\cal S}^{\beta m'_{s_f} m'_{s_i} \ast}_{\lambda (l's')JM}(N'\bfk_f,
N\bfk_i;r') R^{\beta\alpha}_{\lambda\kappa J(l's')(ls)}(r',r;\omega) {\cal S}
^{\alpha m_{s_f} m_{s_i}}_{\kappa (ls)JM}(N'\bfk_f, N\bfk_i;r).
\end{eqnarray}
in place of Eq.~(\ref{eq:Rs}).

\section{Numerical calculations}
\label{sec:num}

We apply the DWIA formalism presented in the previous sections to the 
intermediate energy ($\vec{p},\vec{n}$) reactions around the quasi-elastic 
peaks. We calculate the polarized cross sections $ID_q$ and $ID_p$ of the 
reactions observed at LAMPF\cite{ref:TA94} and RCNP\cite{ref:WA99} as 
summarized in Table~\ref{tab:thq}.

\subsection{Choice of parameters}

First we present the parameters that we selected for our numerical 
calculations.

\subsubsection{Optical potentials}

For the optical potentials of the incident proton and the outgoing neutron, 
we took the global optical potential given by Cooper {\it et al.}\cite
{ref:CO93}. It is designed for the proton but we used it even for the neutron
by excluding the Coulomb part. To study the optical potential dependence, we
also used the proton optical potential given by Jones {\it et al.}\cite
{ref:JO94} and the neutron global optical potential given by Shen {\it et al.}
\cite{ref:SH91}. For the potentials which have the form of the Dirac 
phenomenology, we rewrote them in the Schr\"{o}dinger equivalent form and 
solved the non-relativistic Schr\"{o}dinger equation to obtain the distorted
waves. We used the relativistic kinematics and the reduced energy 
prescription.

\subsubsection{Effective mass, single-particle potentials and spreading width}

To obtain the effective mass, we have to determine $f(r)$ and $U_L(r)$, as was
discussed in Sec.~\ref{sec:kmass}. Suggested by Ref.~\cite{ref:GT83}, we assume
$f(r)$ has the form
\begin{equation}
f(r)=\frac{m^*(r)}{m}=1-\frac{b}{1+\exp\left(\frac{r-R}{a}\right)}
\end{equation}
with $R=r_0 A^{\frac{1}{3}}$, since we consider only the $k$-mass. Note that
$m^*(r=0) \approx (1-b)m$. The parameter $b$ will be adjusted to reproduce the
energy spectra of $ID_q$.

For $U_L$ we used the form
\begin{eqnarray}
U_L(r)&=& -(V_0+\ii W_0)\frac{1}{1+\exp\left(\frac{r-R}{a}\right)} \nonumber
\\
& &-\frac{2}{m_{\pi}^2}\frac{V_{ls}}{a}\frac{\exp\left(\frac{r-R}{a}\right)}
{\left(1+\exp\left(\frac{r-R}{a}\right)\right)^2}{\bf l}\cdot{\bf s}+V_{\rm 
Coul},
\end{eqnarray}
where $V_{\rm Coul}$ is the Coulomb potential of a uniformly charged sphere 
with the radius $R_c=r_cA^{\frac{1}{3}}$. The shape parameters are fixed at 
$r_0=r_c=1.27$ fm and $a=0.67$ fm\cite{ref:BM75}, and the nucleon spin orbit
potential depth $V_{ls}$ is chosen to be 6.5 MeV for $^{12}$C and 10.0 MeV for
$^{40}$Ca, respectively\cite{ref:NI95}. The real potential depth $V_0$ is 
determined in such a way as to give the observed separation energy of the 
outermost occupied state of the target nucleus.

Using the phenomenological energy-dependent relation for the spreading 
width\cite{ref:MN81}
\begin{equation}
\gamma(\epsilon)=21.5\left[\frac{(\epsilon-\epsilon_{\rm F})^2}
{(\epsilon-\epsilon_{\rm F})^2+18^2}\right]\left[\frac{110^2}
{(\epsilon-\epsilon_{\rm F})^2+110^2}\right]
\end{equation}
in units of MeV with the Fermi energy $\epsilon_{\rm F}$, we set the imaginary
potential parameter $W_0$ for the particle to be $W_0=\frac{1}{2}\gamma 
(\epsilon_p)$, and the hole spreading width $\gamma_h$ to be $\gamma_h=\gamma
(\epsilon_h)$.

For $\Delta$, we set $m_{\Delta}=1232$ MeV, $V_0=30$ MeV and $W_0=V_{ls}=0.0$
MeV\cite {ref:NI95}.

\subsubsection{Effective interaction for ring approximation}

We take RPA correlation into account only in the isovector spin-vector channel
($T=1, S=1$ in Sec.~\ref{sec:isodel}). For other channels, we simply use the
uncorrelated response functions. For the effective interaction in the isovector
spin-vector channel, we employ the ($\pi+\rho+g'$) model, in which it is 
written as
\begin{equation}
V^{\alpha\beta}(\bfr_1-\bfr_2;\omega)=(\btau^{\alpha}_1\cdot\btau^{\beta}_2)
\left[ V_{\rm L}^{\alpha\beta}(\bfr_1-\bfr_2;\omega)+V_{\rm T}^{\alpha\beta}
(\bfr_1-\bfr_2;\omega)\right]
\end{equation}
with
\begin{eqnarray}
V_{\rm L}^{\alpha\beta}(\bfr_1-\bfr_2;\omega)&\equiv&\int\frac{d^3q}{(2\pi)^3}
{\rm e}^{\ii \bfq\cdot(\bfr_1-\bfr_2)}W_{\rm L}^{\alpha\beta}(q,\omega)(\bsigma^
{\alpha}_1\cdot\hat{\bfq})(\bsigma^{\beta}_2\cdot\hat{\bfq}), \\
V_{\rm T}^{\alpha\beta}(\bfr_1-\bfr_2;\omega)&\equiv&\int\frac{d^3q}{(2\pi)^3}
{\rm e}^{\ii \bfq\cdot(\bfr_1-\bfr_2)}W_{\rm T}^{\alpha\beta}(q,\omega)(\bsigma^
{\alpha}_1\times\hat{\bfq})\cdot(\bsigma^{\beta}_2\times\hat{\bfq}).
\end{eqnarray}
Here $W_{\rm L}^{\alpha\beta}$ and $W_{\rm T}^{\alpha\beta}$ are given by
\begin{eqnarray}
W_{\rm L}^{\alpha\beta}(q,\omega)&=&\frac{f_{\alpha}f_{\beta}}{m_{\pi}^2}\left
(g'_{\alpha\beta}+\Gamma^{\pi}_{\alpha}(q,\omega)\Gamma^{\pi}_{\beta}
(q,\omega)\frac{q^2}{\omega^2-q^2-m_{\pi}^2}\right), \\
W_{\rm T}^{\alpha\beta}(q,\omega)&=&\frac{f_{\alpha}f_{\beta}}{m_{\pi}^2}\left
(g'_{\alpha\beta}+C^{\rho}_{\alpha\beta}\Gamma^{\rho}_{\alpha}(q,\omega)
\Gamma^{\rho}_{\beta}(q,\omega)\frac{q^2}{\omega^2-q^2-m_{\rho}^2}\right),
\end{eqnarray}
where $m_{\pi}$ ($=139$ MeV) and $m_{\rho}$ ($=770$ MeV) are the pion and the
$\rho$-meson masses, respectively, and the coefficient $C^{\rho}_{\alpha\beta}
$ is the ratio of the $\rho$-meson coupling to the pion one. The 
$q$-dependence of the Landau-Migdal parameters $g'_{\alpha\beta}$ is 
neglected\cite{ref:DM87} and the vertex form factors are chosen to be
\begin{equation}
\Gamma^{\pi}_{\alpha}(q,\omega)=\frac{m_{\pi}^2-\Lambda_{\pi}^2}
{\omega^2-q^2-\Lambda_{\pi}^2},\hspace{3mm} \Gamma^{\rho}_{\alpha}(q,\omega)
=\frac{m_{\rho}^2-\Lambda_{\rho}^2}{\omega^2-q^2-\Lambda_{\rho}^2}
\end{equation}
where $\Lambda_{\pi}$ and $\Lambda_{\rho}$ are the cutoff parameters. We note
that we can identify $\Delta^*$ with $\Delta$ in $g'_{\alpha\beta}$, $f_
{\alpha}$, $C^{\rho}_{\alpha\beta}$ and $\Gamma^{\pi(\rho)}_{\alpha}$.

We set the parameters as $f_N^2=1.0$, $f_{\Delta}/f_N=2.0$, $C^{\rho}_
{\alpha\beta}=2.18$, $\Lambda_{\pi}=1300$ MeV, and $\Lambda_{\rho}=2000$ 
MeV\cite{ref:AEM82}. As to the Landau-Migdal parameters, we fix $g'_{NN}=0.6$,
$g'_{\Delta\Delta}=0.5$ and we adjust $g'_{N\Delta}$ to reproduce the observed
$ID_q$ as well as possible.

\subsubsection{Energy Shift}
The nuclear structure model used here is a simple one which is based on the 
mean field approximation with the RPA correlation, because we are interested
in only gross structure of highly-excited states in the continuum. Apparently
the model does not well represent the structure of the low lying states, 
especially the ground states of the deformed nucleus $^{12}$C. Therefore the
excitation energy obtained by the present model should be smaller than the 
observed one, because the real ground state energy should be lower than that
given by the model.

To remedy this shortcoming, we made the following modification. The observed
energy spectra of $^{12}$C\cite{ref:CHEN93,ref:WA99} show the eminent peak of
the unresolved $4^-$ and $2^-$ states of $^{12}$N. We artificially add 5 MeV
to the energy transfer $\omega$ to coincide the calculated peak to the 
observed one for the uncorrelated case ($m^*=m$ and no RPA correlation). For
the correlated case, this energy shift becomes 3.5 MeV, but we kept the 5 MeV
shift for all different parameter sets. This small difference of energy shift
does not affect our conclusion. Any shift was not made for $^{40}$Ca.

\subsubsection{Convergence}

In the actual calculations, we have to limit the infinite range of the 
summations and integrations by judging from their convergence. The maximum 
angular momenta of the distorted waves were fixed at 40, and the maximum 
transferred angular momenta were set at 9 for $^{12}$C and 12 for $^{40}$Ca 
for both the incident energies, because the transferred momenta are close in
both cases.

The maximum momentum for the Fourier transformation from the momentum space 
to the coordinate space was set to be 4 fm$^{-1}$, though the main 
contribution comes from $q$ near the observed transferred momenta shown in 
Table~\ref {tab:thq}. The radial integration was limited up to 10 fm, since 
the response functions are well damped at this radius.

\subsection{Results of DWIA calculations}

First we try to reproduce the spin-longitudinal cross section $ID_q$ of the 
$^{12} {\rm C}(\vec{p},\vec{n})$ reactions at 494 MeV by adjusting the 
Landau-Migdal parameter $g'_{N\Delta}$ and the effective mass at the center 
$m^*(r=0)$, namely parameter $b$, within a reasonable range.

\begin{figure}[t]
\begin{center}
\epsfbox{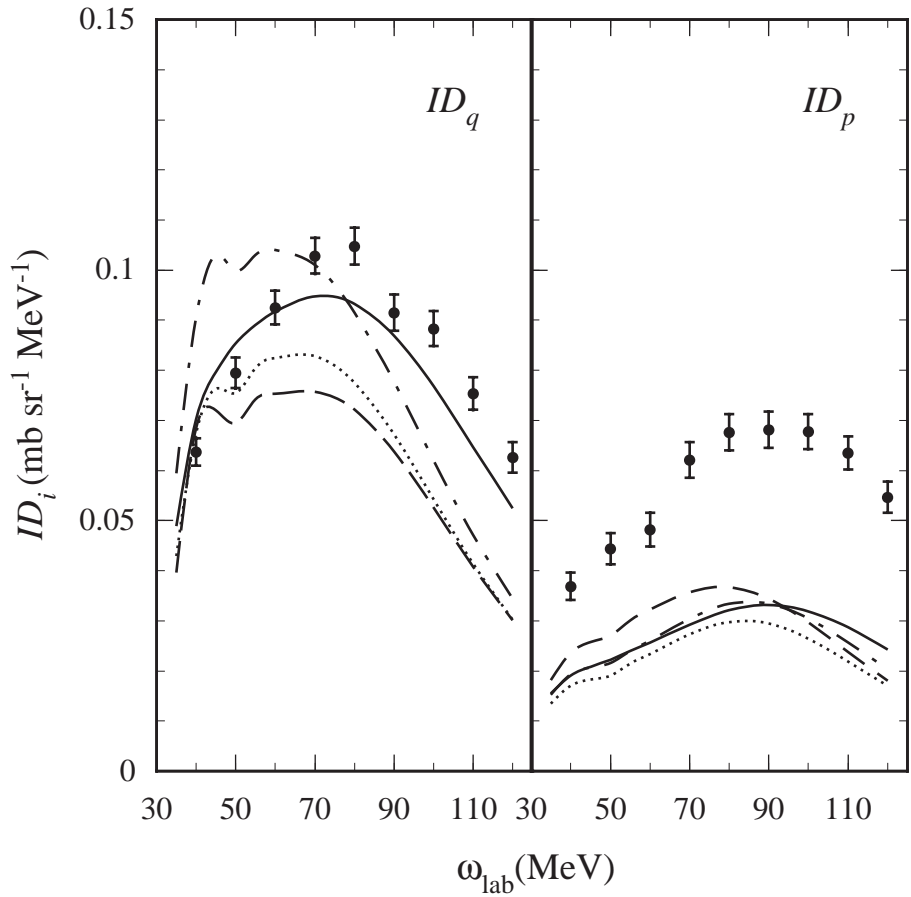}
\end{center}
\vspace{3mm}
\caption{DWIA results of $ID_q$ and $ID_p$ for $^{12}$C($\vec{p},\vec{n}$) at
494 MeV. The dashed line denotes the result with $m^*=m$ and without the RPA
correlation. The dotted and the dot-dashed lines represent the RPA results of
$(g'_{NN}, g'_{N\Delta}, g'_{\Delta\Delta})=(0.6, 0.6, 0.6)$ and $(0.6, 0.3,
0.5)$ with $m^*=m$, respectively. The solid line shows the RPA results of $
(0.6, 0.3, 0.5) $ with $m^*(r=0)=0.7m$. Experimental data are taken from 
Taddeucci {\it et al.}[8]}
\label{fig:xsc500}
\end{figure}

The results are shown in the $ID_q$ part of Fig.~\ref{fig:xsc500}. The dashed
line denotes the result with $m^*(0)=m$ (i.e. $b=0$) and without the RPA 
correlation. The dotted line denotes the RPA results with the universality 
ansatz $g'_{NN}=g'_{N\Delta}=g'_{\Delta\Delta}=0.6$ and again with $m^*(0)=m$.
We see that the experimental data are much larger than these two results, and
therefore we reduced $g'_{N\Delta}$, guided by the fact that the response 
functions increase as $g'_{N\Delta}$ decreases\cite{ref:NI95}. The dot-dashed
line shows the RPA results with $(g'_{NN}, g'_{N\Delta}, g'_ {\Delta\Delta})
=(0.6, 0.3, 0.5)$ and $m^*(0)=m$. Now $ID_q$ drastically increases but the 
peak position is much lower than the observed one, owing to the softening. We
then reduced the effective mass $m^*$, guided by the prediction of the Fermi
gas model ($\omega_{\rm peak}=q^2/2m^*$) and the numerical calculation of 
Itabashi\cite{ref:ITA96}. The full line represents the RPA result with $(g'_
{NN}, g'_{N\Delta}, g'_ {\Delta\Delta})=(0.6, 0.3, 0.5)$ and with $m^*(0)
\approx 0.7m(b=0.3)$. By this change of $m^*$, the peak position moves up very
close to the observed one, though the peak value is reduced. This is 
consistent with the Fermi gas prediction ($3m^*/4q p_{\rm F}$, $p_{\rm F}$ 
being the Fermi momentum). Now the fit is very much improved by the choice of
these parameters for $ID_q$.

The results of $ID_p$ are shown in the right panel of Fig.~\ref{fig:xsc500}.
We see that the RPA correlation with $g'_{NN}=g'_{N\Delta}=g'_{\Delta\Delta}
=0.6$ markedly quenches $ID_p$, as was predicted, whereas the observed data 
are very much enhanced. When we reduce $g'_{N\Delta}$ the results increase 
considerably, but are still quenched at low $\omega$. When we further reduce
the effective mass, the peak shifts upwards. We found that all of the 
calculated results are much smaller than the observed $ID_p$. In the end, we
could not reproduce the observed $ID_p$ by changing these parameters within 
reasonable ranges.

\begin{figure}[t]
\begin{center}
\epsfbox{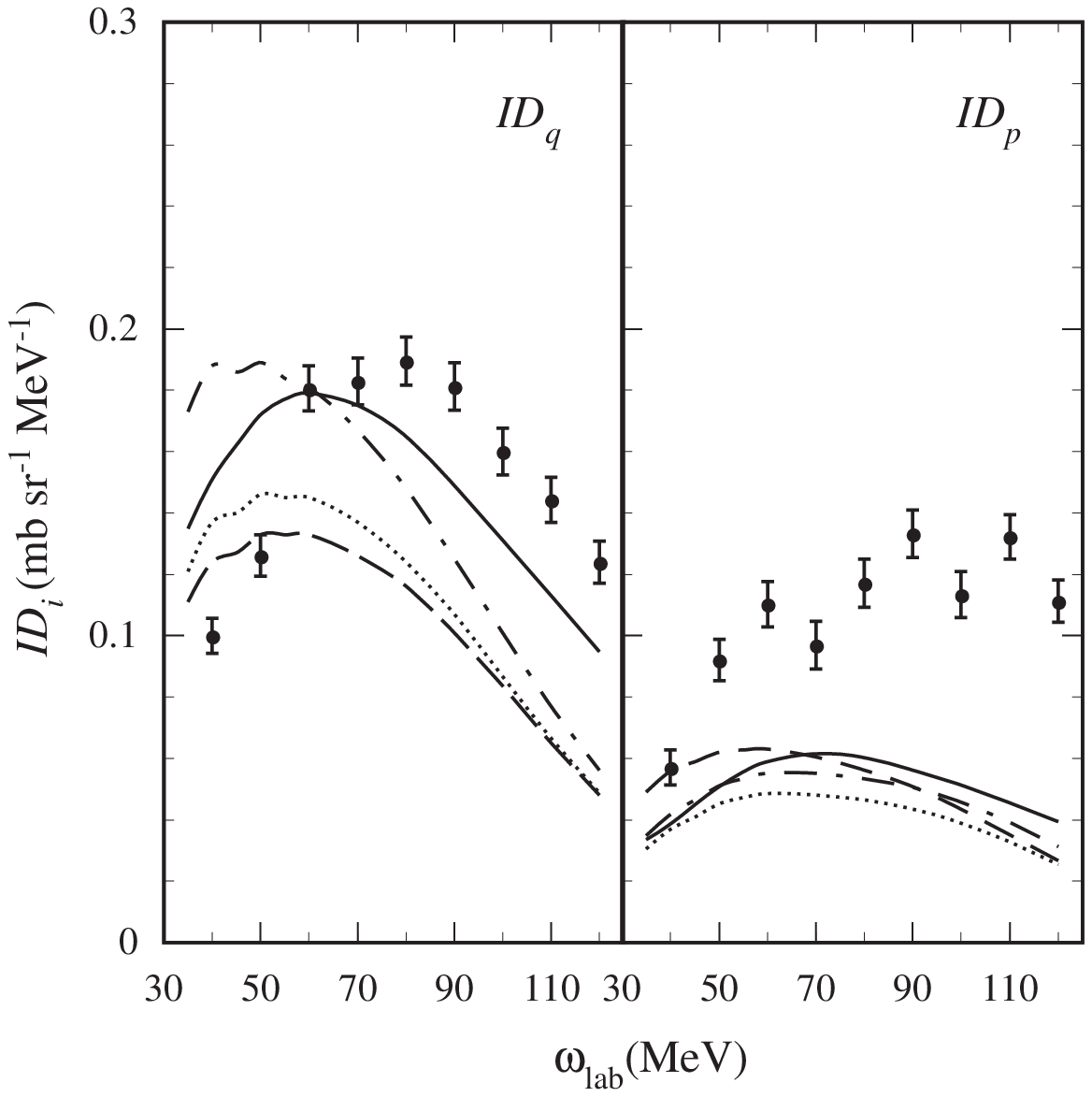}
\end{center}
\vspace{3mm}
\caption{The same as Fig.\ref{fig:xsc500},but for $^{40}$Ca($\vec{p},\vec{n}
$).}
\label{fig:xsca500}
\end{figure}

\begin{figure}[t]
\begin{center}
\epsfysize=8.5cm
\epsfbox{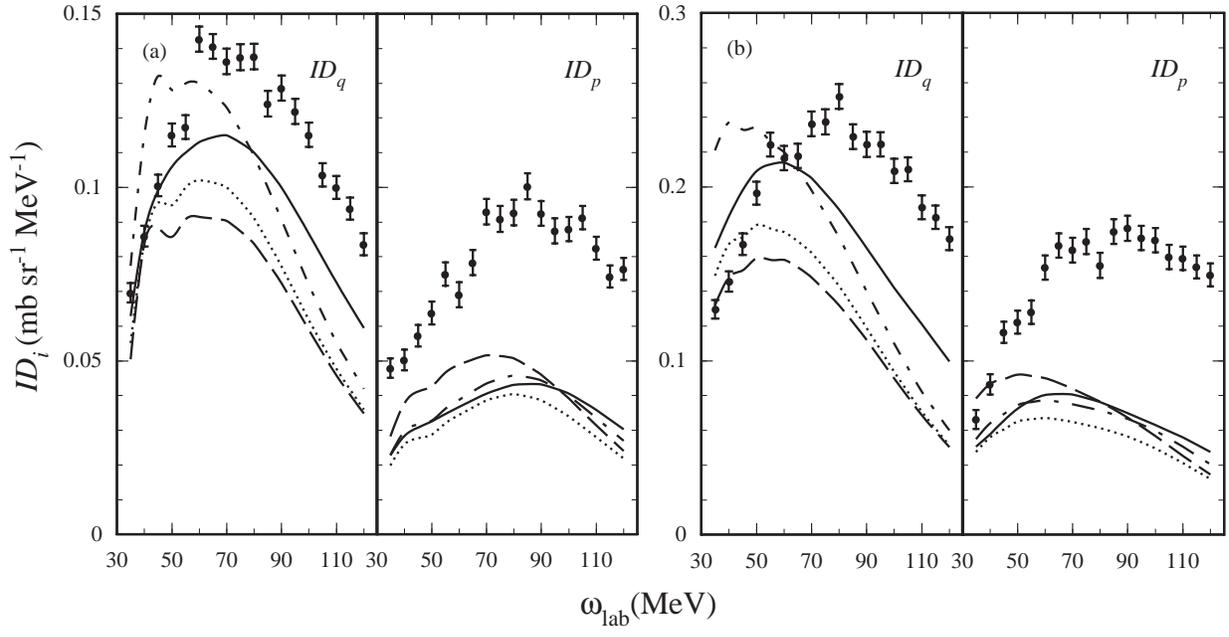}
\end{center}
\vspace{3mm}
\caption{The same as Fig.\ref{fig:xsc500}, but for (a) $^{12}$C($\vec{p},\vec
{n}$) and (b) $^{40}$Ca($\vec{p},\vec{n}$) at 346 MeV. The experimental data
are taken from Wakasa {\it et al.}[9]}
\label{fig:xsc350}
\end{figure}

With the same sets of parameters we calculated the other reactions given in 
Table~\ref{tab:thq}. Fig.~\ref{fig:xsca500} shows the results for $^{40}$Ca$
(\vec{p}, \vec{n})$ at 494 MeV and Fig.~\ref{fig:xsc350} shows the results
for $^{12}$C, $^{40}$Ca $(\vec{p}, \vec{n})$ at 346 MeV. The features of the
results are common for all cases, though the fit of $ID_q$ is somewhat better
for the case of 494 MeV than for that of 346 MeV.

\section{Test of effective nucleon number approximation}
\label{sec:eik}

In the previous experimental papers\cite{ref:TA94,ref:WA99} the response 
functions were extracted by means of PWIA with the effective nucleon number 
approximation described in Sec.~\ref{sec:PWIA}. We test this prescription by
comparing the results with those of DWIA.

For detailed analysis, we also introduce mode-dependent effective nucleon 
numbers defined by\cite{ref:WA99}
\begin{equation}
N_{\rm eff}^{\rm L}(\omega)=N \frac{I^{\rm DW}D^{\rm DW}_q(\omega)}{I^{\rm PW}
D^{\rm PW}_q(\omega)},\ \ 
N_{\rm eff}^{\rm T}(\omega)=N \frac{I^{\rm DW}D^{\rm DW}_p(\omega)}{I^{\rm PW}
D^{\rm PW}_p(\omega)} .
\label{eq:Neff}
\end{equation}
They depend on the spin-longitudinal (L) and spin-transverse (T) modes as well
as the excitation energy $\omega$, though the effective nucleon
number $A_{\rm eff}$ determined by Eq.~(\ref{eq:Aeff}) (effective neutron 
number $N_{\rm eff}$ in the present case) is independent of them.

In the present analysis, we try to get $N_{\rm eff}^{\rm L(T)}(\omega)$ 
without the $\omega$-dependence by estimating them at a certain $\omega$ near
the peak. The obtained $N_ {\rm eff}^{\rm L(T)}$ with and without the RPA
correlation are summarized in Table~\ref{tab:neff}.

For the uncorrelated cases, we found that $N_{\rm eff}^{\rm L}\approx N_{\rm
eff}^{\rm T}$ and thus we set them equal and denote both of them by $N_{\rm 
eff}^{(0)} $ in common. We note that Wakasa {\it et al.} \cite{ref:WA99} 
reported that $N_{\rm eff} $ estimated by Eq.~(\ref{eq:Aeff}) and $N_{\rm eff}
^{(0)}$ obtained above are very close, and also note that the estimation of 
$N_{\rm eff}$ by Eq.~(\ref {eq:Aeff}) is somewhat ambiguous because the 
equation includes the uncertain total $NN$ cross section in the nuclear 
medium. Therefore here we identify $N_{\rm eff}$ with $N_{\rm eff} ^{(0)}$.

\begin{figure}[t]
\begin{center}
\epsfysize=17cm
\epsfbox{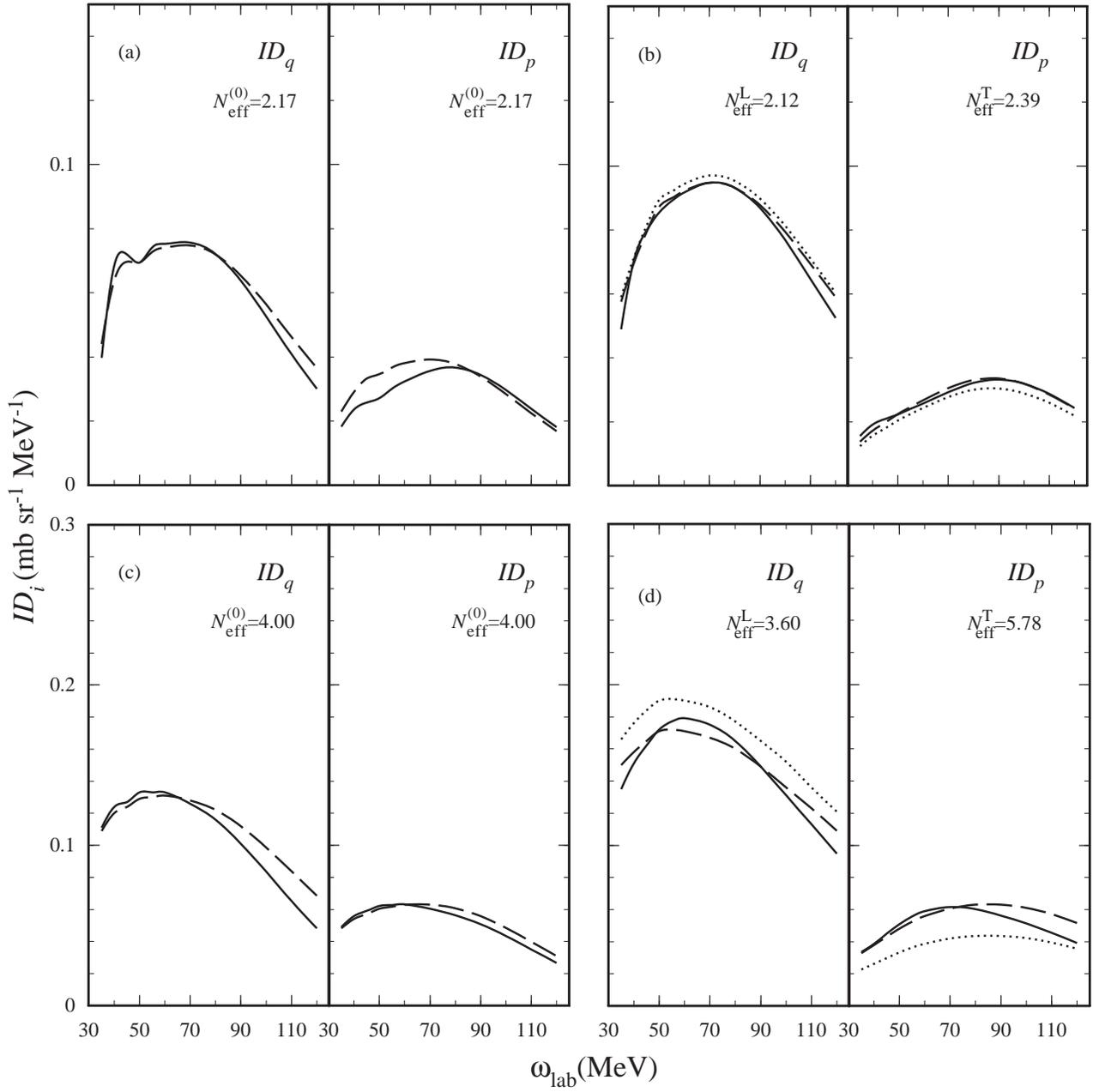}
\end{center}
\vspace{3mm}
\caption{Comparison of DWIA and PWIA results at 494 MeV: (a) without and (b)
with the RPA correlation for $^{12}$C($\vec{p},\vec{n}$) and (c) without and
(d) with the RPA correlation for $^{40}$Ca($\vec{p},\vec{n}$). The solid and
the dashed lines denote the DWIA and the PWIA results multiplied by $N^{(0)}
_{\rm eff}/N(=N_{\rm eff}/N)$ in (a) and (c) and by $N^{\rm L(T)}_{\rm eff}
/N$ in (b) and (d) , respectively. The dotted lines denote the PWIA results 
multiplied by $N_{\rm eff} /N$ in (b) and (d). The values of $N^{(0)}_{\rm 
eff}$ and $N^{\rm L(T)} _{\rm eff}$ are shown in the figure.}
\label{fig:nef500}
\end{figure}

\begin{figure}[t]
\begin{center}
\epsfysize=17cm
\epsfbox{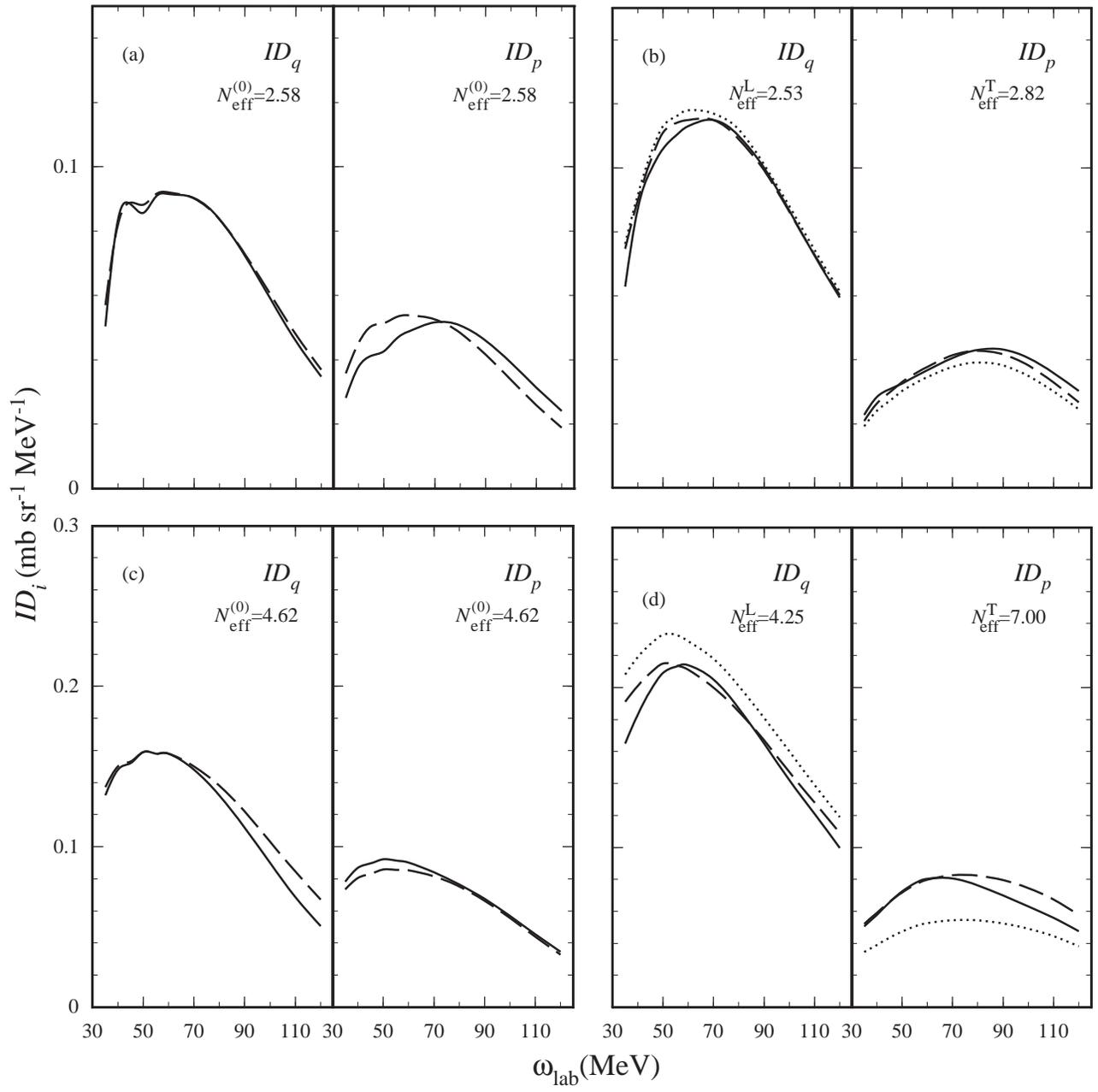}
\end{center}
\vspace{3mm}
\caption{The same as Fig.\ref{fig:nef500}, but at 346 MeV.}
\label{fig:nef350}
\end{figure}

The PWIA cross sections multiplied by $N_{\rm eff}^{0}/N=N_{\rm eff}/N$ 
(dashed lines) and the DWIA cross sections (solid lines) without the 
correlation are compared in Fig.~\ref{fig:nef500}(a) and (c) and Fig.~\ref 
{fig:nef350}(a) and (c) for cases of $^{12}$C and $^{40}$Ca with the incident
energy 494 MeV, and $^{12}$C and $^{40}$Ca with 346 MeV, respectively. The 
$\omega$-independent approximation holds well only for the $ID_q$ of $^{12}$C
at 346MeV, but not so well for other cases. This is an indication of poorness
of the effective nucleon number approximation.

The results with the RPA correlation are compared in Fig.~\ref{fig:nef500}(b)
and (d) and Fig.~\ref{fig:nef350}(b) and (d) for cases of $^{12}$C and $^{40}
$Ca with 494 MeV, and $^{12}$C and $^{40}$Ca with 346 MeV, respectively. The
DWIA results, the PWIA results multiplied by $N_{\rm eff}^ {\rm L(T)}/N$ and
the PWIA results multiplied by $N_{\rm eff}/N$ are denoted by the solid, the
dashed and the dotted lines, respectively. The RPA calculations are carried 
out with $(g'_{NN}, g'_ {N\Delta}, g'_{\Delta\Delta}) =(0.6, 0.3, 0.5) $ and
$m^*(0) =0.7m$. Once the RPA correlation is included, $N_{\rm eff} ^{\rm T} 
$ is increased but $N_{\rm eff}^{\rm L}$ is reduced and thus they differ very
much to each other. Especially the difference is very large for $^{40} $Ca. 
This is another strong indication of poorness of the $N_ {\rm eff}$ method. 
The above changes of $N_ {\rm eff}^{\rm L}$ and $N_{\rm eff} ^{\rm T}$ due to
the RPA correlation may be explained in the following way. The enhancement of
$R_{\rm L}$ and the quenching of $R_{\rm T}$ are stronger in the 
higher-density region (the inner part) but they are masked by the stronger 
absorption in this region. Consequently, we should decrease $N_{\rm eff}^{\rm
L}$ to reduce the enhancement seen in PWIA but increase $N_{\rm eff} ^{\rm T}
$ to reduce the quenching.

The large difference, especially for $^{40}$Ca, between the results of the 
effective nucleon number approximation (the dotted lines) and the DWIA results
(the solid lines) clearly shows that the conventional way of obtaining the 
response functions does not work quantitatively. Therefore the DWIA 
calculation is definitely necessary in the quantitative analysis.

\section{Discussion}
\label{sec:dis}

\subsubsection{Effect of the spin-orbit force}

\begin{figure}[t]
\begin{center}
\epsfysize=8.5cm
\epsfbox{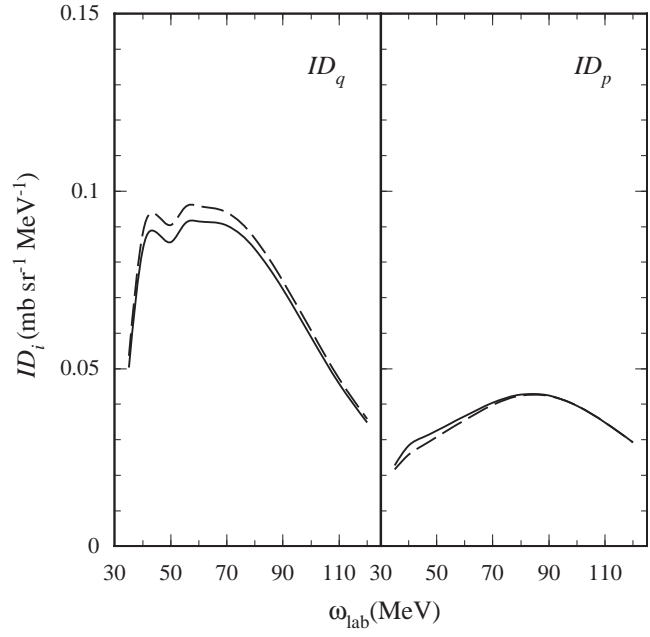}
\end{center}
\vspace{3mm}
\caption{Comparison of the DWIA results with (full line) and without (dashed
line) the spin orbit force for $^{12}$C($\vec{p},\vec{n}$) at 346 MeV.}
\label{fig:ls}
\end{figure}

One of our main concerns in the DWIA calculation is the effects of the spin 
orbit force of the optical potentials. We compared the numerical results with
and without the spin orbit force in Fig.~\ref{fig:ls} for $^{12}$C($\vec{p} 
,\vec{n}$) at 346 MeV. The RPA correlation was not included in this analysis.
We found that effects are larger for $I^{\rm DW}_{\rm lab}D^{\rm DW}_q$ than
for $I^{\rm DW}_{\rm lab}D^{\rm DW}_p$. Fortunately, however, they are so 
small that the spin-orbit force does not greatly disturb the separation 
between the spin-longitudinal and spin-transverse responses. We also reported
similar results for $^{12}$C($\vec{p},\vec{n}$) at 494 MeV\cite{ref:INIK97}.

\subsubsection{Ambiguity of the optical potential}

\begin{figure}[t]
\begin{center}
\epsfysize=8.5cm
\epsfbox{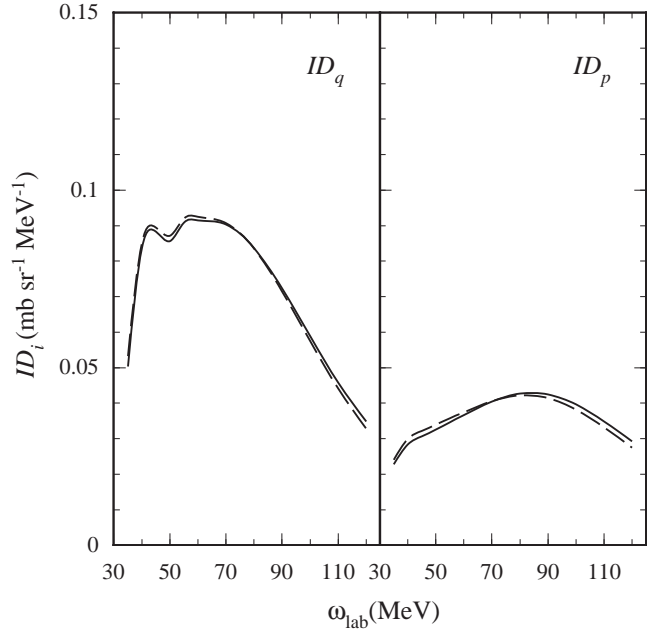}
\end{center}
\vspace{3mm}
\caption{Comparison of the results with various optical potentials for $^{12}
$C($\vec{p},\vec{n}$) at 346  MeV. The full lines denote the results with the
potential obtained by Cooper {\it et al.}[42] for both the proton and the 
neutron, while the dashed lines denote the results with the potential by Jones
{\it et al.}[43] for the proton and by Shen {\it et al.}[44] for the neutron.}
\label{fig:opt}
\end{figure}

To investigate the effects of the ambiguity of the optical potentials, we 
compared the results in terms of the global optical potential given by Cooper
{\it et al.}\cite{ref:CO93} with those in terms of the proton optical 
potential given by Jones {\it et al.}\cite{ref:JO94} and the neutron global 
optical potential given by Shen {\it et al.}\cite{ref:SH91}. Fig.~\ref
{fig:opt} shows the comparison of $I^{\rm DW}_{\rm lab}D^{\rm DW}_q$ and $I^
{\rm DW}_{\rm lab}D^ {\rm DW}_p$ for $^{12}$C($\vec{p},\vec{n}$) at 346 MeV.
Here the RPA correlation was not included. We found that the ambiguity hardly
affects the results.

\subsubsection{Ambiguity of the $NN$ amplitude}

\begin{figure}[t]
\begin{center}
\epsfysize=8.5cm
\epsfbox{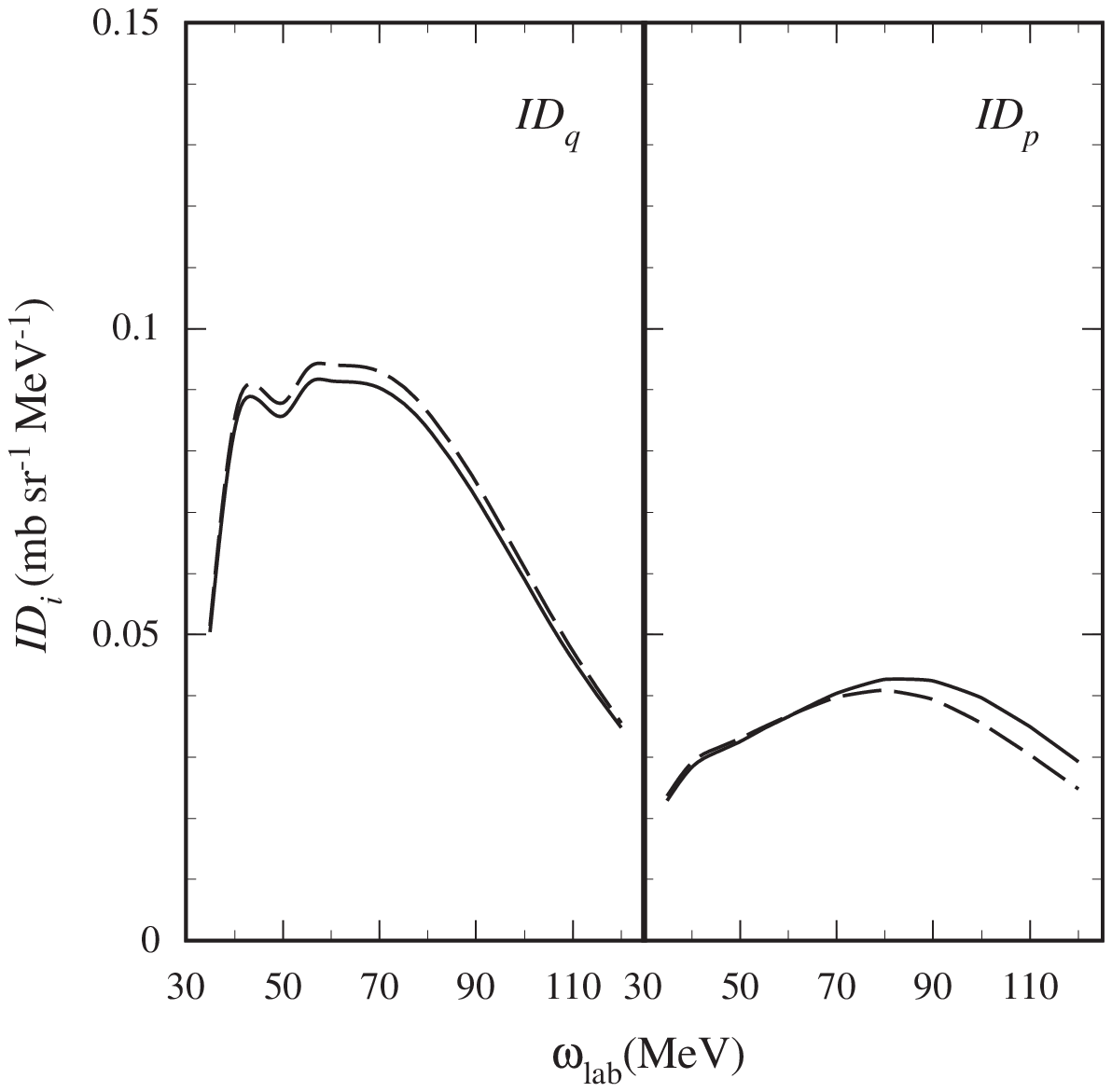}
\end{center}
\vspace{3mm}
\caption{Comparison of the results with the $NN$ amplitudes given by Bugg and
Wilkin (full line)[51] and by Arndt (dashed line)[52] for $^{12}$C($\vec{p}
,\vec{n}$) at 346 MeV.}
\label{fig:amp}
\end{figure}

It has been pointed out\cite{ref:WA99} that the theoretical results are 
affected to some extent by the ambiguity of the $NN$ scattering amplitudes 
obtained by the phase shift analyses. To determine the influence of this 
ambiguity we compared the results with the $NN$ amplitude obtained by Bugg and
Wilkin\cite{ref:BUG92} and with that obtained by Arndt and Roper\cite
{ref:AR94}. Fig.~\ref {fig:amp} shows the results of $I^{\rm DW}_{\rm lab}D^
{\rm DW}_q$ and $I^{\rm DW}_{\rm lab}D^{\rm DW}_p$ for $^{12}$C($\vec{p},\vec
{n}$) at 346 MeV without the RPA correlation.  The effects are less than
10$\%$, and they are different for $I^{\rm DW}_{\rm lab}D^{\rm DW}_q$ and
$I^{\rm DW}_{\rm lab}D^{\rm DW}_p$.

\subsubsection{Two-step processes}

De Pace\cite{ref:DP95} calculated the one- and the two-step contributions in
terms of the Glauber approximation. He found that the two-step process is more
effective for $ID_p$ than for $ID_q$. He concluded that $ID_q$ is reasonably
explained but that there is a large discrepancy between the observed and the
one-step results for $ID_p$, and that the two-step contribution is sizable but
not sufficient to provide an explanation of the large discrepancy. The 
$\Delta$ contribution was not included in his analysis. He made the simple 
assumption for the two-step process that the first step is caused by
the spin-scalar isospin-scalar and the second step by the pure $E$
($F$) amplitude and vice versa.

Nakaoka and Ichimura\cite{ref:NAI99} evaluated the two-step contribution 
through its ratio to the one-step one obtained by PWIA with an on-energy-shell
approximation. They took into account all five terms of the $NN$ amplitudes 
in both the first and second steps. They also found that the two-step process
is more effective for $ID_p$ than for $ID_q$. For $ID_q$ it accounts in large
part for the discrepancy between the DWIA and the experimental results in high
energy transfer regions. They also found that the two-step contribution does
not help to reduce the large discrepancy seen in $ID_p$. Recently, 
Nakaoka\cite{ref:NA00} reported that the two-step effect becomes twice as much
as the previous result, when the off-energy-shell effect is included.

\subsubsection{Remaining problems}

There still remain various other points to be considered, especially keeping
in mind the large discrepancy in $ID_p$. On the structure side, they are (1)
nuclear correlations beyond RPA which has benn evaluated by various 
authors\cite{ref:PA94,ref:FA94,ref:KO98}, \ \ (2) use of genuine RPA\cite
{ref:SH89} instead of the ring approximation, and \ \ (3) self-consistency 
between the mean field and RPA\cite{ref:SSN98,ref:HS00}, etc. On the reaction
side, they include (1) removal of the optimal factorization approximation, \
\ (2) removal of the averaged $\hat{\bf n}$ approximation, \ \ (3) off-shell
effects of the $NN$ $t$-matrix, and \ \ (4) more realistic $NN-N\Delta$ 
transition $t$-matrix, etc.

\section{Summary and Conclusion}
\label{sec:sum}

We first summarized the general formulas for the polarized cross sections, 
$ID_i$, of the $(\vec{N}, \vec{N'})$ scattering to the continuum, which are 
closely related to the nuclear spin response functions. We then reviewed the
PWIA formalism for the intermediate energy $(\vec{N}, \vec{N'})$ reactions, 
treating the Fermi motion by the optimal factorization and the absorption by
the effective nucleon number approximation. The $\Delta$ isobar degrees of 
freedom are also included. This has been widely used to extract the nuclear 
response functions.

Since the reliability of this method has been questioned, we here developed 
the DWIA formalism for the intermediate energy $(\vec{N}, \vec{N'})$ reactions
to the continuum incorporated with the continuum RPA method. This formalism 
is described in the angular momentum representation in the coordinate space 
with the response functions, $R_{J(l's')(ls)} (r',r;\omega)$, which are 
non-diagonal in the coordinate space as well as in the spin space. The 
formalism still involves the optimal factorization and the averaged reaction
normal approximation.

The response function calculation includes the radial-dependent effective mass
and the spreading widths of the particle and the hole. The RPA correlation are
calculated by use of the interaction of the ($\pi + \rho + g'$) model.

The presented formalism is applied to $^{12}$C, $^{40}$Ca $(\vec{p}, \vec {n})
$ at 346 and 494 MeV in the quasi-elastic region. In this analysis, the 
Landau-Migdal parameters and the effective mass are treated as adjustable 
parameters.

From this analysis, we draw the following conclusions:

\noindent (1) The observed spin-longitudinal cross sections $ID_q$ are 
reasonably well reproduced by adjusting $g'_{N\Delta}$ and $m^*$ and by adding
the 2-step contribution. The analysis indicates that the smaller $g'_{N\Delta}
\ (\approx 0.3)$ and the smaller effective mass at the center ($m^*\approx 
0.7$) are preferable. This claim of the smaller $g'_{N\Delta}$ is consistent
with the conclusion obtained from the analysis of the GT 
resonances by Suzuki and Sakai\cite{ref:SS99}.

\noindent (2) The observed spin-transverse cross sections $ID_p$ are much 
larger than the calculated ones and cannot be reproduced within the present 
theoretical framework. However, one must note that $R_T$ corresponding to the
observed $ID_p$ is also much larger than that obtained by $(e,e')$. This 
confirms the previously reported conclusion\cite{ref:TA94,ref:WA99}.

\noindent (3) The $N_{\rm eff}$ method, which has been conventionally used to
obtain the response functions, is found to be a quantitatively poor 
approximation, especially for heavier nuclei such as Ca. The DWIA analysis 
is definitely needed in the quantitative analysis.

\noindent (4) From these findings, we stress that the observation of $R_L/R_T
\le 1$ does not necessarily constitute evidence against the enhancement of 
$R_L$ and of pions in nuclei. We rather conclude that the present experimental
data support the enhancement of $R_L$. The contradiction regarding the ratios
comes from the extraordinarily enhanced $ID_p$. Before drawing any definite 
conclusion about the response functions, we must disentangle the anomalously
large $ID_p$ puzzle.

\noindent (5) We also investigated the effects of the spin-orbit distortion 
and the ambiguity of the optical potential and the $NN$ $t$-matrix, and found
that they are not especially significant.

\acknowledgments

We would like to express our deep gratitude to the late Carl Gaarde and Thomas
Sams with whom we started this project. We are grateful to Terry Taddeucci, 
Jack Rappaport, Hide Sakai and Tomotsugu Wakasa for providing us with 
experimental information and for valuable discussions. This work was supported
in part by the Grants-in-Aid for Scientific Research Nos.~02640215, 05640328
and 12640294 of the Ministry of Education, Science, Sports and Culture of 
Japan.

\vspace{5mm}
\begin{center}
\begin{table}[h]
\caption{Reactions to be analyzed. The values of $q_{\rm cm}$ are those around
the peak.}
\vspace{2mm}
\begin{tabular}{ccccc}
Target\hspace{3mm} & $E_{i,{\rm lab}}$ (MeV)\hspace{3mm} & $\theta_{\rm lab}
$\hspace{3mm} & $q_{\rm cm}$ $({\rm fm}^{-1})$\hspace{3mm} & Refs. \\
\hline
$^{12}$C & 494 & $18^{\circ}$ & 1.70 & \cite{ref:TA94}\\
         & 346 & $22^{\circ}$ & 1.67 & \cite{ref:WA99}\\
\hline
$^{40}$Ca & 494 & $18^{\circ}$ & 1.71 & \cite{ref:TA94}\\
          & 346 & $22^{\circ}$ & 1.68 & \cite{ref:WA99}
\end{tabular}
\label{tab:thq}
\end{table}

\begin{table}[h]
\caption{Effective neutron number. "No" means the no-correlation 
(uncorrelated) cases.}
\vspace{2mm}
\begin{tabular}{ccccc}
$E_i$ (MeV)&Target    &Correlation&$N^{\rm L}_{\rm eff}$&$N^{\rm T}_{\rm eff}
$ \\
\hline

494       & $^{12}$C &    No& 2.17& 2.17\\
          &          &   RPA& 2.12& 2.39\\
          & $^{40}$Ca&    No& 4.00& 4.00\\
          &          &   RPA& 3.60& 5.78\\
\hline
346       & $^{12}$C &    No& 2.58& 2.58\\
          &          &   RPA& 2.53& 2.82\\
          & $^{40}$Ca&    No& 4.62& 4.62\\
          &          &   RPA& 4.25& 7.00
\end{tabular}
\label{tab:neff}
\end{table}
\end{center}

\newpage

\newpage

\newpage

\newpage

\newpage

\newpage

\newpage

\begin{references}
\bibitem{ref:OST92} F. Osterfeld, Rev. Mod. Phys. {\bf 64}, 4922 (1992).
\bibitem{ref:RS94} J. Rapaport and E. Sugarbaker, Annu. Rev. Nuc. Part. Sci.
{\bf 44}, 109 (1994).
\bibitem{ref:AEM82} W. M. Alberico, M. Ericson and A. Molinari, Nucl. Phys. 
{\bf A379}, 429 (1982).
\bibitem{ref:CA84} T. A. Carey {\it et al.}, Phys. Rev. Lett. {\bf 53}, 144 
(1984).
\bibitem{ref:RE86} L. B. Rees {\it et al.}, Phys. Rev. C {\bf34}, 627 (1986)
.
\bibitem{ref:MC92} J. B. McClelland {\it et al.}, Phys. Rev. Lett. {\bf 69},
582 (1992).
\bibitem{ref:CHEN93} X. Y. Chen {\it et al.}, Phys. Rev. C {\bf 47}, 2159 
(1993).
\bibitem{ref:TA94} T. N. Taddeucci {\it et al.}, Phys. Rev. Lett. {\bf 73}, 
3516 (1994).
\bibitem{ref:WA99} T. Wakasa {\it et al.}, Phys. Rev. C {\bf 59}, 3177 (1999)
.
\bibitem{ref:BFS93} G. F. Bertsch, L. Frankfurt, and M. Strikman, Science {\bf
259}, 773 (1993).
\bibitem{ref:BBLW95} G. E. Brown, M. Buballa, Zi Bang Li, and J. Wambach, 
Nucl. Phys. {\bf A593}, 295 (1995).
\bibitem{ref:KO98} D. S. Koltun, Phys. Rev. C {\bf 57}, 1210 (1998)
\bibitem{ref:BS82} G. F. Bertsch and O. Scholten, Phys. Rev. C {\bf 25}, 804
(1982).
\bibitem{ref:ETB85} H. Esbensen, H. Toki, and G. F. Bertsch, Phys. Rev. C {\bf
31}, 1816 (1985).
\bibitem{ref:Al86} W. M. Alberico {\it et al.}, Phys. Rev. C {\bf 34}, 977 
(1986).
\bibitem{ref:Al88} W. M. Alberico {\it et al.}, Phys. Rev. C {\bf 38}, 109 
(1988).
\bibitem{ref:IIKS82} T. Izumoto, M. Ichimura, C. M. Ko, and P. J. Siemens, 
Phys. Lett. {\bf 112B}, 315 (1982).
\bibitem{ref:IKJG89} M. Ichimura, K. Kawahigashi, T. S. J{\o}rgensen and C. 
Gaarde, Phys. Rev. C {\bf 39}, 1446 (1989).
\bibitem{ref:Izu83} T. Izumoto, Nucl. Phys. {\bf A395}, 189 (1983).
\bibitem{ref:PTTW84} A. Picklesimer, P. C. Tandy, R. M. Thaler, and D. H. 
Wolfe, Phys. Rev. C {\bf 30}, 1861 (1984).
\bibitem{ref:Gu86} S. A. Gurvitz, Phys. Rev. C {\bf 33}, 422 (1986).
\bibitem{ref:ZMW87} X. Q. Zhu, N. Mobed, and S. S. M. Wong, Nucl. Phys. {\bf
A466}, 623 (1987).
\bibitem{ref:DP95} A. De Pace, Phys. Rev. Lett. {\bf 75}, 29 (1995).
\bibitem{ref:KKSU00} B. T. Kim, D. P. Knobles, S. A. Stotts, and T. Udagawa,
Phys. Rev. C {\bf 61}, 044611 (2000).
\bibitem{ref:SSA88} T. Shigehara, J. Shimizu and A. Arima, Nucl. Phys. {\bf 
A477}, 583 (1988).
\bibitem{ref:SSIT86} E. Shiino, Y. Saito, M. Ichimura and H. Toki, Phys. Rev.
C {\bf 34}, 1000 (1986).
\bibitem{ref:NI95} K. Nishida and M. Ichimura, Phys. Rev. C {\bf 51}, 269 
(1995).
\bibitem{ref:SS99} T. Suzuki and H. Sakai, Phys. Lett. {\bf B455}, 25 (1999)
\bibitem{ref:BL82} E. Bleszynski, M. Bleszynski, and C. A. Whitten, Jr., Phys.
Rev. C {\bf 26}, 2063 (1982).
\bibitem{ref:IK92} M. Ichimura and K. Kawahigashi, Phys. Rev. C {\bf 45}
, 1822 (1992).
\bibitem{ref:NAI99} Y. Nakaoka and M. Ichimura, Prog. Theor. Phys. {\bf 102}
, 599 (1999)
\bibitem{ref:HA63} R. Hagedorn, {\it Relativistic Kinematics} 
(Benjamin/Cummings Pub., Inc., Readings, Mass., 1963).
\bibitem{ref:FW71} A. L. Fetter and J. D. Walecka, {\it Quantum Theory of Many
Particle System}, (McGraw-Hill, NY, 1971), Chap.~5.
\bibitem{ref:Sat83} G. R. Satchler, {\it Direct Nuclear Reactions} (Clarendon,
Oxford, 1983).
\bibitem{ref:LF81} W. G. Love and M. A. Franey, Phys. Rev. C {\bf 24}, 1073 
(1981).
\bibitem{ref:ADM85} W. M. Alberico, A. De Pace and A. Molinari, Phys. Rev. C
{\bf 31}, 2007 (1985).
\bibitem{ref:SB75} S. Shlomo and G. F. Bertsch, Nucl. Phys. {\bf A243}, 507 
(1975).
\bibitem{ref:KI91} K. Kawahigashi and M. Ichimura, Prog. Theor. Phys. {\bf
85}, 829 (1991).
\bibitem{ref:FP87} S. Fantoni and V. R. Pandharipande, Nucl. Phus. {\bf A473}
, 234 (1987)
\bibitem{ref:ITA96} A. Itabashi, dissertation (Univ. of Tokyo, 1996).
\bibitem{ref:PB62} F. G. Perey and B. Buck, Nucl. Phys. {\bf 32}, 353 (1962)
.
\bibitem{ref:CO93} E.~D.~Cooper, S.~Hama, B.~C.~Clark and R.~L.~Mercer, 
Phys.~Rev.~C {\bf 47}, 297 (1993); S.~Hama, E.~D.~Cooper, B.~C.~Clark and 
R.~L.~Mercer, FORTRAN Program GLOBAL, 1993, \\
{\tt ftp://ftp.physics.ohio-state.edu/tmp/global/global.for}.
\bibitem{ref:JO94} K. W. Jones {\it et al.}, Phys. Rev. C {\bf 50}, 1982 
(1994).
\bibitem{ref:SH91} Shen Qing-biao, Feng Da-chun and Zhuo Yi-zhong, Phys. Rev.
C {\rm 43}, 2773 (1991).
\bibitem{ref:MN81} R. D. Smith and J. Wambach, Phys. Rev. C {\bf 38}, 100 
(1988);C. Mahaux and H. Ng\^{o}, Phys. Lett. {\bf B100} , 285 (1981).
\bibitem{ref:DM87} W. H. Dickhoff and H. M\"{u}ther, Nucl. Phys. {\bf A473},
394 (1987).
\bibitem{ref:BM75} A. Bohr and B. Mottelson, {\it Nuclear Structure} 
(Benjamin, New York, 1975).
\bibitem{ref:GT83} Nguyen van Giai and Pham van Thieu, Phys. Lett. {\bf 126B}
, 421 (1983); C. Mahaux and R. Sartor, Nucl. Phys. {\bf A481}, 381 (1988)
, and references therein.
\bibitem{ref:INIK97} M. Ichimura, K. Nishida, A. Itabashi and K. Kawahigashi,
{\it New Facet of Spin Giant Resonances in Nuclei} (edited by H. Sakai {\it 
et al.} , World Scientific, Singapore, 1997) p.~93.
\bibitem{ref:INIK99} M. Ichimura, K. Nishida, A. Itabashi and K. Kawahigashi,
{\it Highlights of Modern Nuclear Sructure} (edited by Aldo Covello, World 
Scientific, Sigapore, 1999), p.~393.
\bibitem{ref:BUG92} Data provided by D. V. Bugg through T. Sams (1992); D. V.
Bugg and C. Wilkin, Phys. Lett. {\bf 152B}, 37 (1985).
\bibitem{ref:AR94} R. A. Arndt and L. D. Roper, {\it Scattering Analysis 
Interactive Dial-in program} (SAID), Phase-shift solution SP98, Virginia 
Polytechnic Institute and State University (unpublished).
\bibitem{ref:NA00} Y. Nakaoka, Phys. Lett. (submitted), preprint in 
arXiv:nucl-th/0008045.
\bibitem{ref:PA94} V. R. Pandharipande {\it et al.}, Phys. Rev. C {\bf 49}, 
789 (1994).
\bibitem{ref:FA94} Al Fabrocini, Phys. Lett. {\bf B322}, 171 (1994)
\bibitem{ref:SH89} T. Shigehara, K. Shimizu and A. Arima, Nucl. Phys. {\bf 
A492}, 404 (1989).
\bibitem{ref:SSN98} T. Suzuki, H. Sagawa and Nguyen Van Giai, Phys. Rev. C
{\bf 57}, 139 (1998).
\bibitem{ref:HS00} I. Hamamoto and H. Sagawa, Phys. Rev. C {\bf 62}, 024319 
(2000).
\end{references}
\end{document}